\documentclass[acmsmall,screen]{acmart}

\AtBeginDocument{%
  }

\usepackage{amssymb}
\usepackage{amsmath}

\usepackage{dirtytalk}	
\usepackage{xcolor}
\usepackage{multirow}
\usepackage{tikz} 
\usetikzlibrary{positioning,shapes,arrows}
\usepackage{url}
\usepackage{cancel}
\usepackage{subcaption}
\usepackage{float}
\usepackage[normalem]{ulem} 
\usepackage{xspace} 
\usepackage{hyphenat}

\newcommand{\cui}{\texttt{Con\-tin\-ued Use In\-ten\-tion}\xspace}
\newcommand{\CUI}{\texttt{CUI}\xspace}
\newcommand{\pe}{\texttt{Per\-for\-mance Ex\-pec\-tan\-cy}\xspace}
\newcommand{\PE}{\texttt{PE}\xspace}
\newcommand{\ee}{\texttt{Ef\-fort Ex\-pec\-tan\-cy}\xspace}
\newcommand{\EE}{\texttt{EE}\xspace}
\newcommand{\si}{\texttt{So\-cial In\-flu\-ence}\xspace}
\newcommand{\SI}{\texttt{SI}\xspace}
\newcommand{\hm}{\texttt{He\-don\-ic Mo\-ti\-va\-tion}\xspace}
\newcommand{\HM}{\texttt{HM}\xspace}
\newcommand{\fc}{\texttt{Facili\-tating Con\-di\-tions}\xspace}
\newcommand{\FC}{\texttt{FC}\xspace}
\newcommand{\pv}{\texttt{Price Value}\xspace}
\newcommand{\PV}{\texttt{PV}\xspace}
\newcommand{\hab}{\texttt{Habit}\xspace}
\newcommand{\HAB}{\texttt{HT}\xspace}

\begin{document}



\title{From Early Adoption to Sustained Use: Understanding GenAI Usage Among Software Developers in Italian SMEs} 

\author{Fabio Calefato}
\affiliation{%
  \institution{University of Bari}
  \city{Bari}
  \country{Italy}}
\orcid{0000-0003-2654-1588}
\email{fabio.calefato@uniba.it}

\author{Alexandra Pajonk}
\email{alex.pajonk@t-online.de}
\affiliation{%
  \institution{University of Hamburg}
 \city{Hamburg}
  \country{Germany}}

\author{Victoria Jackson}
\email{v.jackson@soton.ac.uk}
\orcid{0000-0002-6326-931X}
\affiliation{%
  \institution{University of Southampton}
 \city{Southampton}
  \country{UK}
}

\author{Guilherme Vaz Pereira}
\email{guilherme.v003@edu.pucrs.br}
\orcid{0009-0006-3521-6081}
\affiliation{%
  \institution{Pontifícia Universidade do Rio Grande do Sul}
 \city{Porto Alegre, RS}
  \country{Brazil}}

\author{Rafael Prikladnicki}
\email{rafaelp@pucrs.br}
\orcid{0000-0003-3351-4916}
\affiliation{%
  \institution{Pontifícia Universidade do Rio Grande do Sul}
 \city{Porto Alegre, RS}
  \country{Brazil}}

\author{Filippo Lanubile}
\affiliation{%
  \institution{University of Bari}
  \city{Bari}
  \country{Italy}}
\orcid{0000-0003-3373-7589}
\email{filippo.lanubile@uniba.it}

\renewcommand{\shortauthors}{Calefato et al.}

\begin{abstract}
Generative AI tools are rapidly transforming software development practice, prompting unprecedented research interest. However, existing studies have predominantly examined initial adoption rather than sustained use. Understanding what drives developers to continue using these tools after initial adoption remains underexplored, particularly in small and medium-sized enterprises where resource constraints shape technology decisions differently than in large organisations.
This study investigates factors associated with developers' intentions to continue using GenAI tools, adapting the UTAUT2 framework to post-adoption professional contexts.
We employed a two-phase mixed-methods design. Phase 1 comprised a six-month longitudinal pilot study at an Italian software company combining surveys and interviews with 17 developers to explore how perceptions of GenAI evolve as experience accumulates. These insights informed a structural model tested in Phase 2 through a cross-sectional survey of 154 developers across Italian SMEs, analysed using PLS-SEM.
The model explained substantial variance in continued use intention ($R^2 = 0.647$), with individual-level perceptions, particularly around productivity, enjoyment, and ease of use, driving sustained adoption, whereas social and organisational factors played no significant role.
These findings suggest that, for GenAI tools, post-adoption behaviour differs from initial adoption patterns: in voluntary professional contexts, sustained use is driven primarily by individual-level factors rather than by social and organisational support.

\end{abstract}



\maketitle



\section{Introduction}
\label{intro}

Generative Artificial Intelligence (GenAI) tools have become pervasive in software development, with organisations adopting them in pursuit of productivity gains and faster time-to-market.
All respondents to the survey conducted by Cycode~\cite{cycode2026state} confirmed having AI-generated code in their codebases and GitHub~\citeyearpar{github_copilot_impact} reports 55\% faster task completion is achieved with Copilot.
Yet, despite the widespread adoption and bold claims, developers remain sceptical.
According to Stack Overflow's 2025 Developer Survey~\cite{2025StackoverflowDevSurvey}, 84\% of developers report using AI tools, with over half using them daily, yet only one-third trust the accuracy of AI-generated output. 
Empirical evidence from industry supports this scepticism. Google's 2024 and 2025 DORA reports~\cite{googleDORA2024,googleDORA2025}  found that delivery stability decreases as AI adoption increases, while GitClear's 2025 report~\cite{gitclear2025} identified rising code duplication and declining refactoring activity.
A survey conducted by Harness~\cite{harness2025} found that most developers spend more time debugging AI-generated code than writing it manually.
Organizations even report 65\% increased security risks since adopting AI coding assistants~\citep{cycode2026state}.
This tension between widespread adoption and quality concerns raises a fundamental question: \textit{What sustains developers' commitment to these tools beyond initial adoption?}

Most existing research has examined initial adoption intentions or early-stage perceptions of GenAI tools.
Studies have documented that developers value productivity gains such as faster task completion and reduced keystrokes~\citep{liang_large-scale_2024}, yet also struggle with understanding and verifying AI-generated output~\citep{vaithilingam2022expectation}.
What remains unclear is whether and how these early experiences translate into sustained use over time.
Initial adoption and continued use are distinct phenomena~\citep{bhattacherjee2001ect}: the former is driven by expectations about future benefits, while the latter reflects experience-based evaluations of whether those expectations were met.
For organisations investing in GenAI tools, understanding what sustains use is arguably more important than understanding what initiates it.

The existing literature presents two notable gaps.
First, while studies have examined initial GenAI acceptance~\citep{russo_navigating_2024} and early usage patterns~\citep{vaz_pereira_exploring_2025}, empirical evidence on what drives continued use after developers have gained substantial experience remains limited.
Second, existing GenAI research has primarily focused on large enterprises.
Whether findings from these studies generalise to small and medium-sized enterprise (SME) settings remains unclear, given that SMEs operate under distinct conditions, such as tighter resource constraints, smaller teams with less formalised structures, and greater autonomy in tool selection~\citep{pino2008sqj,devos2012rethinking,buonanno2005jeim-factors}.

This study addresses both gaps by examining the factors that drive software developers in Italian SMEs to continue using GenAI tools after initial adoption.
A two-phase sequential mixed-methods design was employed.
Phase~1 consisted of a six-month pilot study at Apuliasoft, an Italian software SME, combining surveys, semi-structured interviews, and ethnographic observations to explore developers' GenAI experiences and identify factors relevant to sustained use.
These insights informed an adaptation of the Unified Theory of Acceptance and Use of Technology 2 (UTAUT2)~\citep{venkatesh2012utaut2} to the post-adoption context.
Phase~2 tested this model through a cross-sectional survey of 154 developers across multiple Italian SMEs, analysed using Partial Least Squares Structural Equation Modelling (PLS-SEM).

Our model achieved substantial explanatory power, accounting for 64.7\% of variance in continued use intention.
Perceived productivity and effectiveness gains (\pe) emerged as the dominant predictor, indicating that tangible performance benefits are the primary driver of sustained GenAI use among developers.
Enjoyment and satisfaction derived from using these tools (\hm) also showed a significant positive effect, while perceived ease of use (\ee) demonstrated a smaller but significant contribution.
These results reveal that in voluntary, post-adoption contexts within SMEs, developers' commitment to GenAI tools is sustained not only by the performance benefits they experience but also by the intrinsic enjoyment they derive from use.

This study makes several contributions. 
First, we provide empirical evidence of post-adoption behaviour in the professional GenAI context, moving beyond initial acceptance to examine what sustains use over time among developers.
Second, our six-month longitudinal pilot offers evidence of how developer perceptions evolve from nascent experiences to sustained integration, revealing which factors gain or lose importance as experience accumulates.
Third, we validate a UTAUT2-based model specifically adapted to the voluntary, post-adoption context of SME software development, identifying which theoretical predictors remain salient in this setting.
Fourth, we derive practical recommendations for organisations---particularly SMEs---interested in correctly sustaining GenAI adoption among their development teams.
To facilitate replication and extension, we make our survey instruments, codebook, and analysis scripts publicly available.\footnote{\url{https://figshare.com/s/b60572a4cbb18b54e693}. Individual responses and interview transcripts remain confidential due to their sensitive nature.}

The remainder of this paper is organised as follows.
Sect~\ref{sec:background} reviews related work on GenAI adoption in software engineering and the UTAUT2 framework.
Sect~\ref{sec:framework} presents the research model and hypotheses.
Sect~\ref{sec:methods} details the two-phase mixed-methods design.
Sect~\ref{sec:phase1_results} and \ref{sec:phase2-results} present findings from the pilot study and cross-sectional validation study, respectively.
Sect~\ref{sec:discussion} discusses theoretical and practical implications, followed by limitations and future directions.
Finally, we draw conclusions in Sect~\ref{sec:conclusion}.


\section{Background and Related Work} \label{sec:background}



This section positions our study within three domains of existing research: empirical evidence on GenAI use in software development (Sect~\ref{sec:bg-genai}), technology acceptance frameworks for studying continued use (Sect~\ref{sec:bg-acceptance}), and the distinctive characteristics of SME software development contexts (Sect~\ref{sec:bg-sme}).

\subsection{GenAI Adoption in Software Engineering}\label{sec:bg-genai}

GenAI tools have spread rapidly in software development practice and 
a growing body of empirical work has examined how professional developers use GenAI tools, documenting both usage patterns and lack of trust in output quality.
This tension between widespread adoption and persistent scepticism underscores the need to understand what sustains continued use beyond initial experimentation.

Large-scale surveys have documented established usage patterns. \citet{liang_large-scale_2024} found that AI programming assistants generate nearly one-third of code in professional workflows, with developers primarily using these tools for code completion and generation tasks. 
However, participants also reported significant usability challenges, including difficulty crafting effective prompts and managing the cognitive load of evaluating AI-generated suggestions. 
Yet \citet{vaithilingam2022expectation} found that while GitHub Copilot did not improve task completion time in controlled settings, most participants nonetheless preferred it for daily programming, suggesting that factors beyond raw efficiency shape continued use decisions.

Empirical evaluations have documented both capabilities and limitations. 
\citet{imai2022copilot} found that Copilot generates more code than human pair-programming but requires more deletions, suggesting quality trade-offs developers must navigate. 
\citet{moradidakhel2023copilot} found that Copilot generates solutions for most fundamental tasks but struggles when problems require combining multiple methods.
\citet{mastropaolo2023robustness} demonstrated that semantically equivalent prompts produce different outputs, highlighting consistency issues.

Organisational studies have examined GenAI adoption in professional environments. \citet{weisz2025enterprise} conducted an enterprise deployment study at IBM, finding net productivity increases that were not experienced uniformly across developers. 
Other case studies report that developers primarily use GenAI tools for individual tasks such as code generation and learning, with limited integration into team-level workflows~\citep{kemell_still_2025}.
\citet{zhou2025problems} systematically categorised developer-reported problems from GitHub Issues and Stack Overflow (SO), reporting challenges related to context limitations, output quality, and workflow integration that emerge during sustained use.

Controlled experimental evidence is beginning to emerge. \citet{cui2025effects} report on randomised controlled trials with 4,867 developers over 2--8 months, finding a 26\% increase in completed tasks, with junior developers showing the largest gains.

Despite this growing body of work, most existing research examines initial adoption perceptions or early-stage use rather than the factors that sustain GenAI use over time.

\subsection{Technology Acceptance and Continued Use}\label{sec:bg-acceptance}
Technology acceptance research seeks to understand why individuals adopt or reject new technologies to predict and improve adoption outcomes.
The field has evolved from the Technology Acceptance Model (TAM)~\citep{davis1989tam}, which focused on perceived usefulness and ease of use, through the Unified Theory of Acceptance and Use of Technology (UTAUT)~\citep{venkatesh2003utaut}, which integrated several theoretical models into a more comprehensive framework. 
UTAUT2 \citep{venkatesh2012utaut2} extended the original framework to consumer contexts, substantially improving explanatory power for technology adoption intentions.
UTAUT2 identifies seven predictors of technology use, presented in more detail in Sect~\ref{sec:reseach-model}: \pe, \ee, \si, and \fc (retained from UTAUT), plus \hm, \pv, and \hab.

A critical distinction in this literature separates \textit{initial adoption} from \textit{continued use}. 
\citet{bhattacherjee2001ect} established that post-adoption behaviour follows different dynamics than initial adoption: the cognitive basis shifts from expectations about future benefits to evaluations of realised benefits through actual experience. 
Subsequent research has demonstrated that the factors predicting initial adoption may differ from those sustaining long-term use~\citep{karahanna1999pre-post-adoptioon, limayem2007habit}, with habit and confirmation of expectations playing increasingly important roles over time.

Recent studies have applied technology acceptance frameworks to GenAI adoption among software developers. 
\citet{russo_navigating_2024} proposed a Human-AI Collaboration and Adaptation Framework drawing on TAM and Diffusion of Innovation theory~\citep{rogers2003diffusion}, finding that developers adopt GenAI tools when they fit existing workflows.
\citet{lambiase_investigating_2025} found that habit and performance expectancy were primary adoption drivers. 
However, both studies focused on initial adoption rather than continued use. 
Whether findings from these studies generalise to post-adoption behaviour in resource-constrained SME settings remains an open question that our research addresses. 
Section~\ref{sec:framework} presents our adapted research model and develops hypotheses grounded in UTAUT2 theory.

\subsection{Software Development in SME Contexts}\label{sec:bg-sme}

SMEs represent a critical segment of the global economy and the software industry~\citep{worldBank}. 
In the EU, SMEs account for 99\% of all enterprises, employ approximately 90 million people, and represent the majority of employment in the digital sector~\citep{smeEUAnnualReport}. 
Italy, where this study is situated, has over 3.9 million SMEs~\citep{italy2024AnnualReport}, making them central to the national economy and software sector.

Software SMEs differ systematically from large enterprises in ways that affect how they operate and adopt new technologies. 
SMEs are characterised by limited human and financial resources, smaller teams, constrained technical expertise, and narrower professional networks~\citep{richardsonSmallSoftware2007}. 
\citet{turner2018pm} identified further disadvantages, including restricted access to IT tools and limited technical knowledge.
Additionally, while their small size enables agility and customer responsiveness~\citep{richardsonSmallSoftware2007}, it also constrains the resources available for adopting and sustaining new tools and practices~\citep{pino2008sqj}.

These structural characteristics shape how SMEs approach technology adoption. 
They typically rely on informal peer learning~\citep{coetzer2017distinctive} and grant employees greater autonomy in tool selection due to limited dedicated support~\citep{devos2012rethinking}. 
Compared to larger organisations, they face distinct barriers to AI adoption, including limited infrastructure, lower technology readiness, and persistent skills gaps~\citep{schwaeke2025ainewnormal, ayinaddis2025aiinsme}.

These characteristics have direct implications for GenAI adoption. 
Resource constraints may amplify the importance of perceived performance benefits, given SMEs' limited capacity to sustain tools during long learning curves. 
Limited formal training infrastructure means developers must learn GenAI tools through practice and peer support rather than structured programmes. 
The smaller scale of small teams may increase the visibility of peer practices, potentially amplifying social influence effects. 
Because existing empirical research on GenAI adoption has largely focused on large enterprises~\citep{vaz_pereira_exploring_2025, weisz2025enterprise}, there is limited evidence on whether these findings generalise to SMEs, where organisational dynamics differ substantially.



\section{Theoretical Framework and Hypotheses}\label{sec:framework}

\begin{figure}[tb]
\centering
\includegraphics[width=9cm]{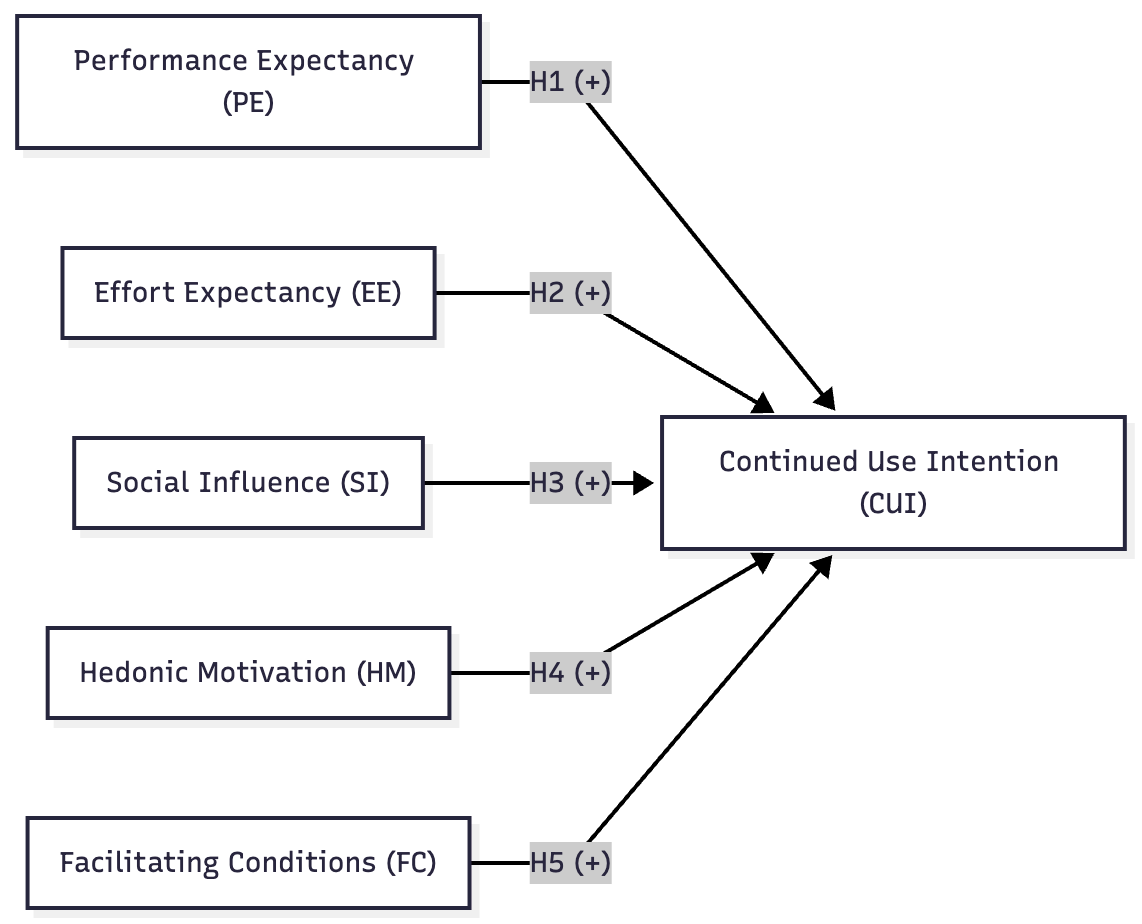}
\caption{UTAUT2-based model of continued GenAI use in software organisations.}
\label{fig:research_model}
\end{figure}

\subsection{Research Model Overview}\label{sec:reseach-model}

We ground our study on UTAUT2 theory presented in Sect.~\ref{sec:bg-acceptance}, adapting it from its original focus on initial adoption intentions to our context of \textit{continued use} among software developers in organizational settings.
This adaptation required a systematic analysis of construct relevance to post-adoption behaviour.

Our adapted model (see Figure~\ref{fig:research_model}) was empirically validated through a six-month longitudinal case study at Apuliasoft, an SME based in Bari, Italy, which tracked developers' perceptions as they evolved from early voluntary use to sustained use of GenAI tools over six months. 
The model retains five core UTAUT2 constructs as predictors of \cui (\CUI): \pe (\PE), the perceived performance benefits of GenAI tools; \ee (\EE), the ease of continued use; \si (\SI), perceived expectations from peers and colleagues; \hm (\HM), the enjoyment derived from use; and \fc (\FC), the availability of organisational and technical support.

The pilot assessed the full UTAUT2 framework with seven constructs. The detailed findings are presented in Sect.~\ref{sec:phase1-pilot}.
Based on theoretical analysis and empirical findings from the triangulated survey, interview, and observational data, two constructs, \hab and \pv, were excluded from the confirmatory Phase~2 model. 
The rationale for these exclusions is presented in Sect.~\ref{sec:phase1-constructs-exclusion}.

\subsection{Construct Definitions and Hypotheses Development}\label{sec:hypotheses}

In this section, we define each construct in the research model (Fig.~\ref{fig:research_model}) and develop the associated hypotheses drawing on UTAUT2 theory, relevant literature, and contextualization to GenAI adoption in SME settings. 

\textbf{\cui (\CUI)} represents developers' plans and commitment to maintain GenAI tool usage in their future work activities. 
Unlike initial adoption intention, which captures users' willingness to begin using a new technology, \CUI reflects an already-established decision to persist with or abandon a technology~\citep{bhattacherjee2001ect}. 
The cognitive basis shifts from expectations about future benefits (pre-adoption) to evaluations of realized benefits through actual experience (post-adoption). 
As the outcome variable in our model, \CUI represents developers' psychological commitment to persist with GenAI tools. This intention, in turn, serves as a proxy for sustained usage behaviour---i.e., whether developers actually continue using GenAI in practice. 
Previous research on post-adoption technology use~\citep{bhattacherjee2001ect} showed that users' stated intentions to continue using a technology strongly predict whether they actually do.

\textbf{\pe (\PE)} represents the degree to which developers believe that continued use of GenAI tools improves their job performance \citep{venkatesh2012utaut2}, encompassing perceptions of enhanced productivity, faster task completion, improved code quality, and better problem-solving capabilities. 
In post-adoption contexts, performance expectancy shifts from expectations about potential future benefits to evaluations of benefits already experienced.

\PE is one of the strongest predictors in technology adoption research~\citep{davis1989tam, venkatesh2003utaut, venkatesh2012utaut2, kuttimani2021utaut2meta}. 
For continued use specifically, Expectation-Confirmation Theory~\citep{bhattacherjee2001ect} posits that post-adoption behaviour is driven primarily by confirmation of performance expectations through actual experience. 
In SE contexts, \citet{lambiase_investigating_2025} have identified perceived productivity gains as a primary driver of GenAI tool adoption. 
The importance of \PE may be amplified in SME settings, where resource 
constraints~\citep{richardsonSmallSoftware2007} make productivity gains 
particularly valuable~\citep{kwarteng2024utautsmeeu}.

\begin{quote}
\textit{H1: Performance expectancy is positively associated with continued use intention.}
\end{quote}

\textbf{\ee (\EE)} captures the degree of ease associated with continued use of tools~\citep{venkatesh2012utaut2}. 
In the case of GenAI tools, it encompasses the cognitive effort required to formulate effective prompts, interpret AI-generated outputs, and integrate GenAI capabilities into existing development workflows. 
In the post-adoption phase, \EE shifts from perceptions of initial learning difficulty to evaluations of ongoing ease after the main learning curve is overcome.

While \EE represents a fundamental dimension of technology acceptance \citep{davis1989tam}, 
its effect on behavioural intention appears to diminish with experience~\citep{venkatesh2003utaut}, 
and recent meta-analytic evidence suggests the path relationship may be weaker than 
originally theorised~\citep{kuttimani2021utaut2meta}.
Qualitative studies of GenAI adoption suggest that perceived ease of use becomes less salient once developers overcome initial learning curves~\citep{russo_navigating_2024}. 
This is particularly relevant in SME settings where formal training 
resources are typically limited, and tools must consequently be learned 
through practice and peer support~\citep{cardon2004managing, coetzer2017distinctive}.
Still, given the post-adoption context of our study, we anticipate \EE will have a weaker influence on \cui compared to \pe.

\begin{quote}
\textit{H2: Effort expectancy is positively associated with continued use intention.}
\end{quote}

\textbf{\si (\SI)} reflects how individuals' behaviour is shaped by their social environment~\citep{venkatesh2012utaut2}. In this context, it represents the extent to which developers perceive that important others (e.g., peers, colleagues, team leaders) believe they should continue using GenAI tools.

Most empirical evidence on \SI derives from initial adoption contexts. Meta-analytic evidence indicates moderate effects overall, with stronger effects in mandatory use settings~\citep{kuttimani2021utaut2meta}, while research on developer tool adoption has found that peer visibility facilitates initial uptake~\citep{witscheyAdoptionSecurityTools2015}.
Whether these effects persist into post-adoption phases remains less clear. 
Recent research on GenAI adoption among software engineers found that social factors did not significantly predict intention to use, suggesting that individual evaluations of utility may outweigh social pressures once developers gain direct experience with the technology~\citep{russo_navigating_2024}.
Despite this uncertainty, the smaller, tightly connected nature of SME development teams could plausibly amplify social influence effects by increasing the visibility of peer practices and fostering more informal, person-to-person knowledge sharing.
We therefore include \SI in our model while acknowledging that its relevance in post-adoption contexts warrants empirical examination.

\begin{quote}
\textit{H3: Social influence is positively associated with continued use intention.}
\end{quote}

\textbf{\hm (\HM)} captures the fun, enjoyment, and intrinsic pleasure derived from using technology~\citep{venkatesh2012utaut2}, reflecting intrinsic motivation, whereby individuals perform activities for their inherent satisfaction rather than for external rewards.
In the GenAI context, this encompasses the enjoyment of creative problem-solving through prompt engineering, the satisfaction of discovering new tool capabilities, and the pleasure of collaborating with an AI assistant using natural language.

UTAUT2 incorporates \HM in consumer contexts~\citep{venkatesh2012utaut2}, recognizing its importance for voluntary use and technologies with creative elements, as well as for continued engagement once initial novelty wanes. 
Meta-analyses found \HM to be significantly associated with intention to use in most studies, though effects are weaker in utilitarian contexts~\citep{kuttimani2021utaut2meta}.
Even in professional software development, typically framed as utilitarian, intrinsic motivation plays a meaningful role: developers value technical challenges and variety in their work~\citep{beecham2008motivation}, and they experience flow states during coding that sustain engagement~\citep{meyer2014productivity}. 
GenAI's interactive, conversational nature fosters creative experimentation that may be intrinsically rewarding. 
In SME settings, where technology mandates tend to be less formalised and developers have greater freedom over tool selection~\citep{devos2012rethinking,buonanno2005jeim-factors}, enjoyment could play an amplified role in sustaining use.

\begin{quote}
\textit{H4: Hedonic motivation is positively associated with continued use intention.}
\end{quote}

\textbf{\fc (\FC)} represent developers' perceptions of the organisational and technical resources and support available to facilitate continued GenAI use~\citep{venkatesh2012utaut2}, encompassing the availability of necessary resources, possession of required knowledge and skills, compatibility with existing tools and workflows, and access to help when difficulties arise.

In post-adoption contexts, \FC may shape \cui: if developers perceive inadequate organisational support, tool incompatibilities, or a lack of resources, they may deem continued use infeasible and abandon GenAI despite recognising its benefits.
The construct may be especially relevant in SME environments, where limited IT budgets, smaller support teams, and less formal training infrastructure make organisational support particularly salient~\citep{cardon2004managing, coetzer2017distinctive}.

\begin{quote}
\textit{H5: Facilitating conditions are positively associated with continued use intention.}
\end{quote}

\section{Research Methodology} \label{sec:methods}



\subsection{Design} \label{design}
This study employs a two-phase sequential exploratory-confirmatory mixed-methods design~\citep{creswell2017} to investigate factors influencing continued GenAI use among software developers in Italian SMEs.
The design combines a longitudinal case study, providing contextual insights, with a broader cross-sectional survey, enabling statistical generalization. 
This approach allows us to first understand the phenomenon deeply within a specific organizational context before testing our theoretical model across multiple settings.
It also aligns with recommendations for context-sensitive technology adoption research~\citep{venkatesh2016review}, ensuring our model is both grounded in the lived experiences of Italian SME developers and validated with adequate statistical power.

\textbf{Phase 1: Pilot Case Study}. 
Phase~1 examines developers’ experiences with GenAI tools over six months at Apuliasoft, an Italian software SME with about 45 developers. 
It adopts a longitudinal, within-subject design to capture how perceptions and usage patterns evolve, offering insights into factors that support or hinder continued engagement with GenAI.

Rather than directly applying the UTAUT2 framework, we first conducted an exploratory mixed-methods pilot study combining surveys, semi-structured interviews, and ethnographic observations of actual GenAI use. 
This approach generated contextual evidence on how developers in Italian SMEs experience and adopt these tools, informing the adaptation of UTAUT2 constructs to this setting.
Following~\citet{creswell2017}’s sequential exploratory design, we triangulated quantitative and qualitative data to identify the constructs most relevant to developers’ continued use intentions. The resulting insights guided the development of the UTAUT2-based structural model described in Sect.~\ref{sec:reseach-model}.

The longitudinal design serves three purposes: (1) grounding the model in empirical observations from the SME context, (2) refining survey measures before broader deployment, and (3) strengthening validity through methodological triangulation. 
The temporal dimension also allows comparison between early expectations and later experiences, revealing how perceptions mature with sustained use, an aspect that cross-sectional designs cannot capture.
A key feature of our design is that it tracks the evolution from early to continued voluntary use, rather than from pre- to post-adoption. This focus highlights the factors that motivate developers to continue using GenAI after initial experimentation, offering insights into how organisations can realize lasting value from these tools.

\textbf{Phase 2: Cross-sectional Validation Study}.
Building on insights from Phase~1, Phase~2 extends the investigation to multiple Italian SMEs through a cross-sectional survey design.
This phase tests the refined theoretical model with adequate statistical power to confirm hypothesized relationships and assess their generalizability beyond the single-case context. 
By recruiting software developers from various Italian SMEs, we examine whether the factors identified in Phase~1 hold across different organizational settings while maintaining focus on the distinctive SME context.

The cross-sectional approach complements the longitudinal depth of Phase~1 by providing breadth of evidence. 
While Phase~1 offers a rich understanding of how continued use intentions develop over time within one organization, Phase~2 assesses whether these patterns extend to the broader population of Italian SME developers.
We employed Partial Least Squares Structural Equation Modeling (PLS-SEM) to validate the structural relationships identified in Phase~1, examining which UTAUT2 factors are most relevant in Italian SME contexts. 

\begin{figure}
    \centering
    \includegraphics[width=1\linewidth]{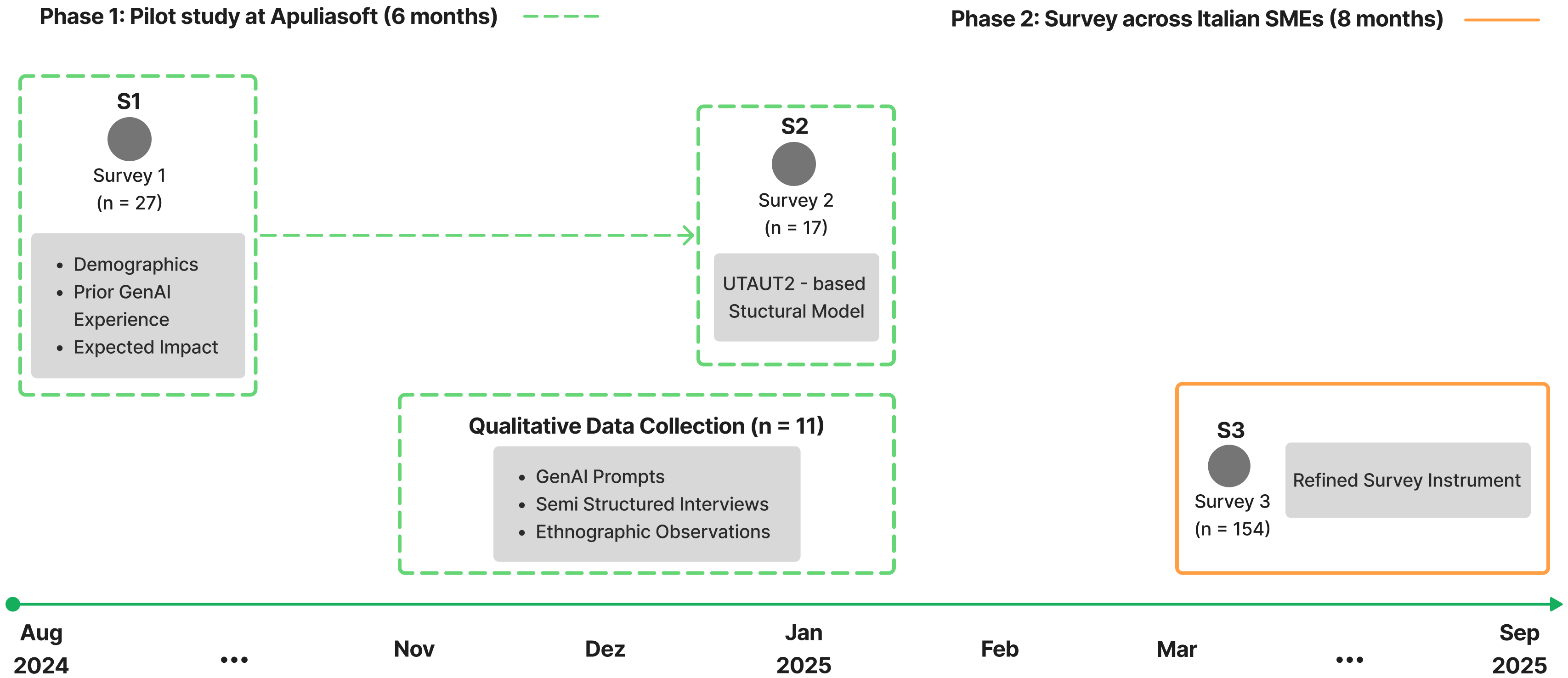}
    \caption{Research design timeline showing Phase 1 (pilot study) and Phase 2 (cross-sectional survey) data collection points.}
    \label{fig:timeline}
\end{figure}

Fig.~\ref{fig:timeline} illustrates the overall timeline of both phases and the data collection steps taken at each point.
Detailed descriptions of each research phase follow in next sections.

\subsection{Phase 1: Pilot Case Study}\label{sec:phase1-pilot}

We selected Apuliasoft Srl, a young and dynamic software development company based in Bari, Italy, as our case study site.
Apuliasoft had 45 employees at the beginning of the study and fit our target company profile as an Italian SME. 
The company's service portfolio includes product development, software maintenance, and consulting services across web and mobile development, data visualization, UX/UI design, and cloud solutions.
Apuliasoft offers flexible work arrangements, allowing staff to choose between remote and office-based work.
One of the researchers involved in the study had prior familiarity with Apuliasoft's organizational culture and employees through a previous internship, facilitating smoother interactions with software engineers and management throughout the study period.
This prior relationship facilitated access to the longitudinal data collection required for the six-month pilot study.

Critically for our study, Apuliasoft provided an ideal voluntary usage context. At the start of the study, the company had no formal guidelines or policies regarding GenAI use, maintaining a deliberately relaxed attitude toward usage. 
In contrast to some companies that mandate GenAI usage (e.g., Microsoft~\citep{microsoftMandateAI}), Apuliasoft management neither promoted nor required the use of GenAI tools. Decisions regarding whether to use GenAI, which tools to select, and how to integrate them into their workflows are left entirely to individuals. 
This authentic voluntary usage environment, combined with management's strategic interest in understanding GenAI's potential through empirical research, created optimal conditions for studying sustained use in professional SME contexts.

\subsubsection{Initial Use Survey (S1)}

We administered an initial survey (S1) in July 2024 to establish baseline perceptions at the start of the study.
The survey reused the questions from an existing instrument validated in a study of GenAI adoption within a Brazilian firm~\citep{pereira_early_2024}, with two additional items about work location to account for Apuliasoft's flexible arrangements. 
The survey covered four main topics (Table~\ref{tab:pre_survey}): demographics (10 items), prior GenAI experience (2 items), GenAI usage at work (6 items), and non-use patterns for those without experience (2 items). 
Key questions addressed what tools developers used, their familiarity with these tools, and their expectations regarding GenAI's impact on development speed, quality, and collaboration.

\begin{table}
    \small
    \centering    
    \caption{Initial survey (S1). The table shows the topics included in the survey, along per topic, the number of questions, and a subset of questions.}
    \begin{tabular}{p{0.18\textwidth}p{0.1\textwidth}p{0.6\textwidth}}
    \hline
        \textbf{Topic} & \textbf{Num}. & \textbf{Questions (Subset)}\\
    \hline
        Demographics & 10 & Age, Gender, Years of experience, Job Role, Primary Work location, Days in office\\
        Prior Gen AI experience & 2 & Experience, Usage at Work \\
        GenAI Usage At Work & 6 & Tools used, familiarity with use, how they believe GenAI influences development speed and adoption of best development practices,  impact on team collaboration and communication, concerns about usage\\
        Non-use & 2 & Reasons for not using, potential interest in using GenAI in the future\\
    \hline
    \end{tabular}
    \label{tab:pre_survey}
\end{table}



The survey instrument was prepared in English, translated into Italian by the researchers from the University of Bari, and administered via Google Forms. 
The company management reviewed the translation to ensure clarity and contextual relevance. 
The CEO distributed the survey to 39 developers through the company's internal communication system, with a reminder email one week later. 
The survey remained open for two weeks. Participation was entirely voluntary, and all responses were anonymous. 
27 developers responded (69\% response rate), labelled P1 through P27.



\subsubsection{Interviews} \label{interviews}

Twelve semi-structured interviews~\citep{seaman1999qualitative} were conducted between July and August 2024 with developers who volunteered to discuss their GenAI experiences in response to the initial use survey. 
The interview protocol covered motivations for using GenAI, trust and reliability perceptions, impact on development work, prompt engineering practices, training needs, and negative experiences. 
While the protocol provided structure, interviewers maintained flexibility to adapt questions based on individual responses in the interview and in their earlier survey. 
The protocol evolved iteratively as interviews progressed, incorporating emerging themes.

All interviews were conducted in Italian by the same researcher, recorded, and automatically transcribed using Whisper AI. 
Transcripts were subsequently reviewed and manually corrected. 
Interview durations averaged 34 minutes (range: 24-42 minutes). Table~\ref{tab:fup-interview-demos} presents participant demographics. 
The 12 interviewees represented diverse career levels and GenAI tool preferences, with ChatGPT being universally adopted and several participants experimenting with additional tools like Copilot, Gemini, and AI image generators.


\begin{table}
    \small
    \centering
    \caption{Interview participant demographics. The ID corresponds to the one assigned in the initial us survey (S1)). Key: Gen.- Gender, Exp.- Years of Professional Experience, Lvl.- Career Level (PD-Professional Developer, AD-Associate Developer, TD-Trainee Developer, PLD-Professional Lead Developer), Spec.- Specialization. AI Image Generators include Dall-E, Stable Diffusion, Midjourney, Leonardo, and Firefly.}
    \begin{tabular}{p{0.04\textwidth}p{0.08\textwidth}p{0.05\textwidth}p{0.06\textwidth}p{0.05\textwidth}p{0.08\textwidth}p{0.42\textwidth}}
    \hline
        \textbf{ID} & \textbf{Age} & \textbf{Gen.} & \textbf{Exp.} & \textbf{Lvl.} & \textbf{Spec.} & \textbf{GenAI Tools} \\
        \hline
        P1 & 26-35 & M & 4-7 & PD & UX & ChatGPT, AI image generators \\
        P2 & 18-25 & M & 1-3 & AD & UX,QA & ChatGPT, Google Gemini   \\
        P5 & 36-45 &F  & 1-3 & AD &  & ChatGPT   \\
        P8 & 26-35 & M & 1-3 & TD &  &ChatGPT   \\
        P11 & 18-25 & M & 1-3 & PLD &  & ChatGPT   \\
        P12 & 26-35 & F & 4-7 & AD &  & ChatGPT  \\
        P13 & 26-35 & M & 1-3 & AD &  & ChatGPT  \\
        P16 & 18-25 & M & 1-3 & AD &  & ChatGPT, Copilot, Gemini, AI image generators  \\
        P19 & 36-45 & M & 4-7 & PD &  & ChatGPT  \\
        P20 & 26-35 & M & 4-7 & PD & & ChatGPT \\
        P25 & 26-35 & M & 4-7 & PLD & UX & ChatGPT \\
        P26 & 18-25 & M & 4-7 & AD & & ChatGPT \\
        \hline
    \end{tabular}
    \label{tab:fup-interview-demos}
\end{table}

\subsubsection{Ethnographic Fieldwork} \label{ethnographic_fieldwork}

Ethnographic fieldwork took place during November and December 2024, coinciding with the middle phase of sustained use. We purposefully decided to undertake the fieldwork two months later than the interviews to allow developers time to embed GenAI tools into their workflows. We initially planned traditional participant observation with the researcher spending time alongside developers observing their daily tasks. 
However, Apuliasoft's flexible remote work arrangements made this operationally challenging. 
We therefore adapted the approach to combine scheduled individual sessions with opportunistic observations.

The scheduled sessions provided structured opportunities for collaborative review of developers' GenAI interactions.
Each session began with feedback validation from earlier interviews, followed by an exploration of participants' chat histories in tools such as ChatGPT or Gemini. 
These sessions offered insights into actual usage patterns, prompt evolution, and integration strategies. 
Eleven such sessions occurred, involving ten interview participants and one additional developer who had completed only the entry survey.

These scheduled sessions complemented more traditional observational work, in which the researcher documented informal conversations with developers and management, capturing spontaneous reflections on GenAI experiences as they emerged naturally in the workplace.


\subsubsection{Sustained Use Survey (S2)}

\begin{table}
    \small
    \centering  
    \caption{Sustained use survey (S2). Topics along with the number of questions and a subset of questions within each topic.}
    \begin{tabular}{p{0.18\textwidth}p{0.1\textwidth}p{0.61\textwidth}}
    \hline
        \textbf{Topic} & \textbf{Num}. & \textbf{Questions}\\
    \hline
        Demographics & 4 & Age, gender, years of experience, job role\\
        GenAI experience & 12 & Familiarity with use; experienced impact on development speed, software quality, and team collaboration; specific use cases\\
        UTAUT2 constructs & 11 & Voluntariness of use, Technology openness, Performance expectancy, Effort expectancy, Social influence, Facilitating conditions, Hedonic motivation, Continued use intention\\
    \hline
    \end{tabular}
    \label{tab:post_survey}
\end{table}

The second survey was administered in February 2025, six months after the initial survey S1, to capture perceptions following sustained engagement. 
The survey instrument incorporated three main components (Table~\ref{tab:post_survey}): demographics (4 items), GenAI experience and impacts (12 items), and UTAUT2 constructs (11 items). 
The experience section adapted items from a validated post-survey used in an earlier longitudinal study~\cite{vaz_pereira_exploring_2025}, enabling cross-context comparison between large enterprises and SMEs. 
The UTAUT2-related items measured the same constructs from the research model presented in Sect.~\ref{sec:reseach-model}.
(see Table~\ref{tab:survey_items} in~\ref{app:phase1} for complete item wordings).

The preparation and distribution of the second survey followed the same protocol as S1.
Participation remained voluntary and anonymous, with no mechanism to match S1 and S2 respondents.
Seventeen participants completed the sustained use survey, labelled P28-P44. 

\subsubsection{Post-Study Interviews}

Following the completion of both study phases, we interviewed the management of Apuliasoft (CEO, COO, and CTO) approximately one year after our initial data collection.
The semi-structured session lasted for 1.5 hours and presented the main quantitative results from Phase~2 alongside patterns observed during the Phase~1 pilot, inviting management to reflect on whether findings resonated with their experience, what organisational changes had occurred since the study period, considering the rapid evolution of the GenAI landscape, and what advice they would offer to other SMEs sustaining GenAI adoption.
This ``voice from the trenches'' approach ensured that our implications are grounded in both empirical findings and lived experience, directly informing the practical recommendations presented in Sect.~\ref{sec:apuliasoft-mgmt-tranches}.

\subsubsection{Data Analysis}

\paragraph{Survey Analysis}

For the initial survey (S1), one researcher calculated descriptive statistics for closed-ended responses and read, summarized, and discussed open-ended responses with two other researchers. 
The sustained use survey (S2) received similar treatment, with descriptive statistics providing quantitative profiles of developer perceptions after six months of sustained use. 
For open-ended questions, responses were read, summarized, and discussed among the same three researchers to identify themes.

While S2 measured UTAUT2 constructs, the small sample size (n=17) precluded any hypothesis testing. 
Therefore, these data served three purposes: (1) validating that UTAUT2 constructs were relevant and measurable in the Apuliasoft context, (2) identifying potential measurement issues to refine for Phase~2, and (3) providing preliminary descriptive evidence of construct relationships to inform hypothesis development. 
Formal hypothesis testing via PLS-SEM occurred in Phase~2 with adequate statistical power.

The anonymous survey design prevented formal paired analysis of S1--S2 changes at the individual level. 
However, the demographic similarity between cohorts enabled meaningful comparison of population-level perception shifts during sustained use (see Table \ref{tab:demographics}).

\begin{table}
\small
\centering
\caption{Demographic characteristics of participants in the initial use survey (S1) and sustained use survey (S2), including age, gender, and professional experience.}
\begin{tabular}{llll}
\hline
\textbf{Attribute}           & \textbf{Answer}                                                         & \textbf{\begin{tabular}[c]{@{}l@{}}Survey (S1)\\ n = 27\end{tabular}} & \textbf{\begin{tabular}[c]{@{}l@{}}Survey (S2)\\ n = 17\end{tabular}} \\ \hline
\multirow{3}{*}{Age}        & 18--25                                                                 & 6 (22\%)                 & 2 (12\%)                  \\
                            & 26--35                                                                 & 17 (63\%)                & 11 (65\%)                 \\
                            & 36--45                                                                 & 4 (15\%)                 & 4 (24\%)                  \\ \hline
\multirow{3}{*}{Gender}     & Male                                                                   & 22 (81\%)                & 14 (82\%)                 \\
                            & Female                                                                 & 4 (15\%)                 & 3 (18\%)                  \\
                            & \begin{tabular}[c]{@{}l@{}}Not informed\end{tabular}      & 1 (4\%)                  & --                        \\ \hline
\multirow{4}{*}{\begin{tabular}[c]{@{}l@{}}Experience in \\ software development\end{tabular}} & \begin{tabular}[c]{@{}l@{}}Less than 1 year \end{tabular} & 2 (7\%)                  & 1 (6\%)                   \\
                            & 1--3 years                                                             & 14 (52\%)                & 8 (47\%)                  \\
                            & 4--7 years                                                             & 11 (41\%)                & 6 (35\%)                  \\
                            & 8+ years                                                               & --                       & 2 (12\%)     \\ 
                            \hline
\end{tabular}
\label{tab:demographics}
\end{table}

\paragraph{Interview Transcripts} 
Qualitative analysis of interview transcripts followed an inductive open coding approach(~\cite{corbin2014basics}).
The researcher who conducted the interviews leveraged their familiarity with the data to identify emergent codes grounded in participants' experiences. Each code received a textual description defining its scope, boundaries, and covered topics. 
Codes evolved iteratively as additional transcripts were analyzed.

After completing the initial coding of all transcripts, one researcher created a preliminary coding scheme by organizing codes into thematic categories. 
A second researcher independently coded a subset of segments, enabling inter-rater reliability assessment. 
Initial agreement reached 87.6\% across 89 compared segments. 
Discrepancies prompted refinement discussions that involved a third researcher, acting as arbiter, and resulted in code merging, splitting, category restructuring, and identification of new categories.

A second comparison round achieved 90.1\% agreement across 170 segments (37\% of all coded data). 
Following discussions to resolve remaining discrepancies, the lead researcher verified the refined scheme against all coded segments and adjusted assignments where necessary.

Finally, a third round was needed to refine the initially identified code of \say{common limitations} used for 17 segments into three more specific codes. This was reviewed with 100\% agreement.

The final coding scheme consisted of 29 codes organized into six categories. Table~\ref{tab:codingscheme} presents a partial view; the complete codebook is available in supplementary materials. This paper reports only on the categories of benefits, challenges, and concerns as these are most relevant to sustained use.

\begin{table}
\small
\centering
\caption{Partial coding scheme showing the Challenges category. Support indicates the number of interview transcript segments assigned to each code.}
\begin{tabular}{l|l|l}
\hline
\textbf{Category}                  & \textbf{Code}        & \textbf{Support} \\
\hline
\multirow{5}{*}{Challenges} & Complex and context specific tasks & 20               \\
                                   & Inaccurate responses  & 5               \\
                                   & Unhelpful for specialized technologies     & 4               \\
                                   & Verbose responses   & 6               \\
                                   & Lack of alternative answers    & 1                \\
                                        
                                   \hline
\end{tabular}
    \label{tab:codingscheme}
\end{table}



\paragraph{Triangulation} 

After analyzing each data source, we systematically triangulated findings across surveys, interviews, and ethnographic observations. 
This triangulation process involved comparing quantitative patterns from survey responses with qualitative themes from interviews and contextual insights from field observations. 
This approach provided a rich, nuanced understanding of continued GenAI use at Apuliasoft, capturing both population-level trends and individual experiences that informed our theoretical model specification for Phase~2.

\subsection{Phase 2: Cross-sectional Validation Study}\label{sec:phase2-main}

\subsubsection{Survey Instrument}\label{sec:survey-instrument}

The Phase~2 survey instrument (S3) built directly upon the Phase~1 sustained use survey (S2) administered at Apuliasoft, retaining the validated UTAUT2 measurement items while incorporating refinements based on pilot findings.
The survey maintained the same three-section structure.


Phase~1 findings informed several refinements to item wording to improve clarity and contextual relevance for Italian software developers; these adjustments are detailed in Sect.~\ref{sec:res-phase1-survey-refinements} alongside the pilot data that motivated them.
In addition, from Phase~1 survey instruments, we retained the screening question in the demographics section asking whether respondents had experience using GenAI tools for software development tasks.
Respondents indicating no experience were directed to a brief exit question about barriers preventing adoption, after which the survey concluded.
This ensured our analytical sample comprised only developers with prior use experience, which is necessary for evaluating continued use intentions.
To identify inattentive respondents in the larger Phase~2 sample, we also embedded an attention check item midway through the UTAUT2 section, instructing participants to select a specific response option~\citep{oppenheimer2009instructional}.
This quality control measure was unnecessary in Phase~1's smaller, organisationally focused context but essential for broader cross-SME recruitment. 
Additionally, items were randomised within construct blocks to reduce order effects and response bias. 
The survey was administered via Google Forms.


\subsubsection{Data Collection}\label{sec:phase2-data-collection}

The target population comprised software developers employed at Italian SMEs, operationalised as companies with 10--250 employees following European Commission definitions.\footnote{\url{https://single-market-economy.ec.europa.eu/smes/sme-fundamentals/sme-definition_en}} 
Participants required current employment in software development roles that involve writing code of any type and across any part of the development process (e.g., application logic, tests, infrastructure), and experience using GenAI tools in professional software development work. 
We defined GenAI tools broadly to include conversational AI assistants (e.g., ChatGPT, Claude, Gemini), code completion tools (e.g., GitHub Copilot), and other AI-powered development aids. 
This inclusive definition was intended to capture the diversity of GenAI tool adoption patterns observed in Phase~1.

Data collection occurred between March and September 2025.
While we sought demographic diversity across experience levels, gender, and technology stacks to strengthen the representativeness of our sample, we faced the practical recruitment constraints of industry surveys commonly reported in prior research~\citep{baltes2022sampling}. 
Although a fully representative sample was not feasible, we employed a convenience sampling approach through multiple channels to maximise coverage of Italian SME software developers.
First, we directly contacted developers at Italian SMEs through the researchers' professional networks. 
Second, we publicised the survey on the researchers' LinkedIn profiles in Italian, targeting posts to software development communities and relevant professional groups. 
Third, we distributed the survey to alumni of the University of Bari's Department of Informatics working in software development roles at Italian SMEs.

Ethical approval and informed consent procedures for both study phases are detailed in~\ref{app:ethics}.

\paragraph{Data Screening}
The Phase~2 survey (S3) collected 186 responses.
We applied sequential filtering to ensure data quality and alignment with the study's focus on GenAI adopters.
First, we excluded seven respondents who reported never having used GenAI tools in a professional context.
Of the remaining 179 responses, we excluded two for failing the attention check, 
17 non-programmers,  
five due to knowledge-usage inconsistency (claiming not to know what an LLM is while simultaneously reporting six or more months of GenAI use), and one for experience contradiction (reporting less than one year of software development experience while claiming more than one year of GenAI usage).
Examination for straight-lining (identical responses across all items) identified no such cases.
The final analytical sample comprised 154 valid responses from software developers who actively use GenAI tools in their professional work.
Missing data were minimal, affecting only 1.02\% of construct item responses, and thus no imputation was required.

\paragraph{Sample Size Determination}
We conducted an \textit{a priori} power analysis using G*Power 3.1~\citep{faul2009gpower} to determine the minimum sample size required for PLS-SEM analysis.
Following recommendations for PLS-SEM research~\citep{russo_pls-sem_2021}, we used the F-test for multiple linear regression as an acceptable approximation, given that the PLS-SEM algorithm estimates path coefficients through ordinary least squares regression~\citep{hair2019plssem}.
With five predictors pointing to the endogenous construct (\CUI), we specified a medium effect size ($f^2 = 0.15$), Type I error probability $\alpha = 0.05$, and statistical power $(1 - \beta) = 0.95$.
These parameters yielded a minimum required sample size of 138 respondents.
Our final analytical sample of 154 valid responses exceeds both the calculated minimum and the  less stringent ``ten-times rule'' threshold of 50 observations, i.e., ten times the maximum number of structural paths directed at any construct~\citep{hair2019plssem}.

\subsubsection{Data Analysis}\label{sec:phase2-analysis}


\paragraph{Analytical Approach}
We employed Partial Least Squares Structural Equation Modeling (PLS-SEM) to test our theoretical model.
Analysis was conducted using SmartPLS 4.1~\citep{ringle2024smartpls}.
PLS-SEM distinguishes between measurement models and structural models. 
The measurement model assesses relationships between unobservable latent constructs (e.g., \pe) and their observable indicators (e.g., individual survey items measuring that construct).
Our model uses reflective measurement, where the latent construct causes observed responses to its survey items~\citep{hair2019plssem}. 
The structural model then examines relationships among these latent constructs, i.e., how the five predictor constructs (\PE, \EE, \SI, \FC, and \HM) influence the outcome construct (\CUI). 
We note that while the analysis plan included all five predictors, measurement issues with \SI led to its exclusion from the final structural model, as detailed in Sect.~\ref{sec:phase2-efa}.
PLS-SEM was selected for its suitability in theory development contexts and its capacity to handle complex models with multiple interconnected constructs~\citep{hair2019plssem,russo_pls-sem_2021}.
This method has gained increasing adoption in empirical SE research and is particularly appropriate for exploratory investigations of emerging technologies~\citep{russo_navigating_2024}. 

\paragraph{Preliminary Scale Validation}
Because several UTAUT2 scales were adapted to the GenAI sustained use context in Italian SMEs, we conducted Exploratory Factor Analysis (EFA) as a preliminary validation step before proceeding to PLS-SEM.
EFA served to verify that adapted items loaded onto their intended constructs and to identify any problematic items requiring removal.
We assessed sampling adequacy using the Kaiser-Meyer-Olkin (KMO) measure (threshold $\geq 0.60$) and Bartlett's test of sphericity ($p < .05$).
Factor extraction followed parallel analysis, retaining factors with eigenvalues exceeding the corresponding average random eigenvalues, corroborated by scree plot inspection.
We applied oblique (oblimin) rotation, as correlated factors were expected.
Items were evaluated against multiple retention criteria: factor loadings $\geq 0.50$ (preferred) or $0.40$--$0.50$ (marginal), cross-loadings $< 0.30$, uniqueness $< 0.60$, and theoretical appropriateness.
As all data were collected via a self-report survey at a single time point, we also assessed potential common method bias using Harman's single-factor test; a single factor explaining less than 50\% of variance suggests common method bias is not a major concern.

\paragraph{Measurement Model Assessment}
Following established PLS-SEM guidelines \citep{hair2019plssem, russo_pls-sem_2021}, we assessed the measurement model through internal consistency reliability, convergent validity, and discriminant validity.
Internal consistency reliability was evaluated using Cronbach's alpha ($\alpha$) and composite reliability (CR), with values $\geq 0.70$ considered acceptable.
CR is preferred in PLS-SEM because Cronbach's alpha tends to underestimate reliability by assuming equal indicator loadings~\citep{hair2019plssem}.
Convergent validity was assessed through indicator outer loadings and Average Variance Extracted (AVE).
Outer loadings $\geq 0.70$ indicate that indicators share at least 50\% of variance with their construct; loadings between $0.40$ and $0.70$ may be retained if their removal does not improve AVE and they hold theoretical importance.
AVE values $\geq 0.50$ confirm adequate convergent validity.
Discriminant validity was evaluated using the Heterotrait-Monotrait (HTMT) ratio of correlations (threshold $< 0.85$ or $< 0.90$)~\citep{henseler2015htmt} and the Fornell-Larcker criterion~\citep{fornell1981evaluating}, which requires that the square root of each construct's AVE exceeds its correlations with other constructs.

\paragraph{Structural Model Assessment}
Structural model assessment followed established PLS-SEM procedures~\citep{hair2019plssem, russo_pls-sem_2021}.
We first examined collinearity among predictor constructs using Variance Inflation Factors (VIF), with values below 3 considered acceptable.
Structural path coefficients were estimated using nonparametric bootstrapping with 5,000 subsamples to assess significance, generating $t$-statistics, $p$-values, and 95\% bias-corrected confidence intervals for each hypothesised relationship.
Explanatory power was assessed using the coefficient of determination ($R^{2}$) and adjusted $R^{2}$; in organisational and technology adoption contexts, $R^{2}$ values of 0.25, 0.50, and 0.75 are typically considered weak, moderate, and substantial, respectively~\citep{hair2019plssem}.
Effect sizes ($f^{2}$) were calculated to assess each predictor's individual contribution to explained variance, with values of 0.02, 0.15, and 0.35 representing small, medium, and large effects.
Predictive relevance was evaluated using PLSpredict with 10-fold cross-validation and 10 repetitions~\citep{shmueli2019predictive}; $Q^{2}_{\text{predict}}$ values greater than zero indicate that the model has predictive capability for out-of-sample observations.
Overall model fit was assessed using the Standardised Root Mean Square Residual (SRMR), which quantifies how well the estimated model reproduces the relationships observed in the data~\citep{hu1999cutoff}.
Lower SRMR values indicate a better fit, with values below 0.10 acceptable.



\section{Results of Phase 1: Pilot Case Study}\label{sec:phase1_results}

This section presents findings from the longitudinal case study at Apuliasoft, which served to (1) characterise how developers integrate GenAI into their workflows, (2) explore which UTAUT2 constructs prove most relevant in a voluntary SME context, and (3) refine both the theoretical model and survey instrument for Phase~2. 
We report quantitative findings from two surveys conducted six months apart---the initial use survey (S1, August 2024) and the sustained use survey (S2, February 2025)---supplemented by qualitative evidence from semi-structured interviews and ethnographic observations of actual GenAI use in practice.

\subsection{Sample Overview} \label{sec:sa}
As described in Sect.~\ref{sec:methods}, survey S1 achieved responses from 27 developers (69\% response rate), while survey S2, conducted six months later, retained 17 participants (63\% retention). 
Table~\ref{tab:demographics} presents the demographic comparison between surveys, showing that the S2 sample remained broadly representative of the original cohort in terms of age, gender, and experience distribution.

All 27 participants in S1 reported prior experience with GenAI tools at baseline, confirming that Apuliasoft represented an environment where GenAI adoption had already occurred organically. 
Of these, 22 (81\%) had used GenAI tools in actual work projects, while 5 (19\%) had experimented with GenAI but not yet applied it professionally. 
ChatGPT was the most widely used tool (81\%), followed by GitHub Copilot (26\%). This baseline confirms that our study captures continued use dynamics rather than initial adoption decisions.

The 37\% attrition rate (10 participants) between S1 and S2 warrants consideration.
The anonymous survey design precluded a within-subject analysis.
However, the aggregate demographics (Table~\ref{tab:demographics}) suggest the retained participants were broadly representative. 
The attrition likely reflects normal survey fatigue rather than systematic selection effects related to GenAI attitudes, though we acknowledge this as a limitation and interpret longitudinal patterns cautiously.

\subsection{Usage Patterns and Expectations Over Time}\label{sec:res-phase1-usage}
We first present findings from each survey, then compare how perceptions shifted over six months of continued GenAI use.

\subsubsection{Results from Initial Survey (S1)}


\paragraph{Impact of GenAI Usage on Development Speed and Software Quality} Notably, all the participants already using GenAI for work projects (22 - 81\%) agreed that the development process speeds up. Among participants using GenAI for work, 68\% believed GenAI use can influence the adoption of better software development practices. 


\paragraph{Impact of GenAI Usage on Communication and Collaboration}
Mixed sentiments were expressed, with some participants feeling that team discussions and collaborative practices could be improved, while others noted concerns about the negative effects of GenAI on team dynamics. Discussions with colleagues were more effective as they could \say{\textit{explore an unfamiliar topic with ChatGPT/Copilot}}(P12) before the discussion. However, concerns on negative effects were voiced, including reduced peer interaction if developers \say{\textit{prefer asking AI rather than a more experienced colleague}} (P4) and worries that GenAI might lead to \say{\textit{less communication in the team}} and loss of \say{\textit{that human confrontation and knowledge exchange}} (P17). 



\paragraph{Reasons for non-use} The five participants who did not use GenAI at work provided several reasons related to low performance and efficiency expectations. For example, P8 felt that GenAI does not always produce correct responses and instead prefers Stack Overflow, as it allows them to consider multiple viewpoints before deciding which approach to take. 
In terms of efficiency, P6 noted that \say{\textit{I have never found the debugging/time-saving ratio of AI-generated code to be advantageous}}. 


\subsubsection{Results from Sustained Use Survey (S2)}
The sustained use survey continued to explore the themes of the first survey, such as the influence of GenAI on speed, code quality, and collaboration. Additionally, as the developers had been using GenAI for six months, they were asked for reflections on additional areas, such as their overall experience and influence on their workflow, as well as the types of tasks they were using GenAI to assist with.

\begin{figure}
\centering
\includegraphics[width=0.7\textwidth]{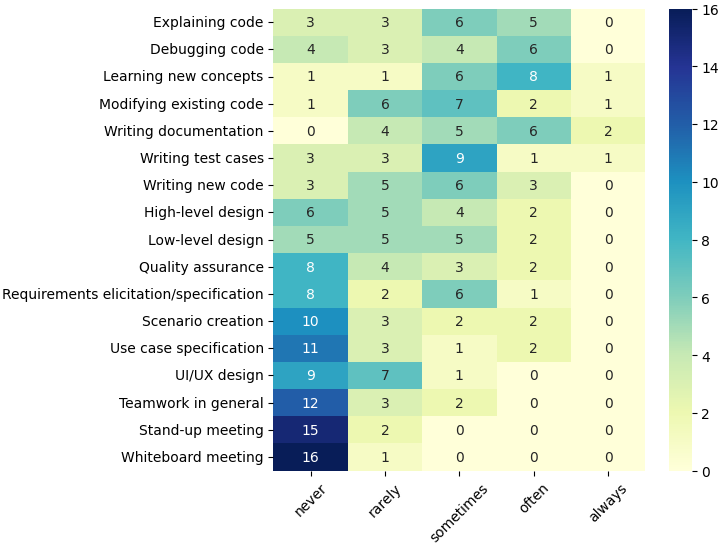}
\vspace{-3mm}
\caption{Types of tasks and GenAI usage frequency (Numbers refer to the number of responses).}
\label{fig:tasks}
\end{figure}

\paragraph{Development Tasks in Which GenAI is Used}
As shown in Figure~\ref{fig:tasks}, GenAI is commonly used to assist with code-related activities, such as explaining code, debugging, modifying existing code, or writing new code. GenAI is also used to help with requirements and design. Low usage was indicated for collaborative tasks, such as stand-ups and whiteboard meetings. This usage pattern aligns with observations from studies exploring GenAI adoption (\citep{vaz_pereira_exploring_2025, davila_industry_2024}).

\paragraph{Usage Frequency}
Looking at how frequently participants use the tools, it is clear that they are well-adopted in developers' daily workflows. Only 12\% indicated using the tools occasionally, while 47\% use them at least several times per week, and 41\% daily. This highlights that developers use GenAI tools habitually.

\paragraph{Overall Experience With GenAI Tools}
Respondents answered that they are generally satisfied (76\%) or very satisfied (12\%); only two respondents had neither a satisfactory nor an unsatisfactory experience (12\%). Even more positive are the results regarding whether GenAI tools have integrated well into participants’ personal workflows, with 36\% responding \say{very well} and 53\% indicating that they have integrated well.



\paragraph{Impact of GenAI Usage on Development Speed and Software Quality}
Looking at the influence that developers perceive GenAI has on development process speed and software quality, it is possible to see why developers continue to use GenAI frequently. The perception is that GenAI has a positive influence on both factors, though not to a great extent. The development process speed is influenced positively, with more than half of respondents (59\%) stating that speed improves significantly or even in a transformative way. No respondent perceives a decline in development speed.

Regarding the impact on software quality, opinions differed slightly. While there is still an overall positive influence, 71\% perceive at least a slight improvement, there are also voices who believe that software quality slightly worsens with the use of GenAI.
Similarly, when asked about the quality of the suggestions provided by the tools, most participants rated it as medium, yet still had a positive impression of the quality. When asked whether GenAI tools are reliable for supporting development activities, responses are rather negative, with suggestions being moderately to somewhat reliable. Nonetheless, respondents indicate that the tools make their tasks easier.


\paragraph{Impact of GenAI Usage on Communication and Collaboration}
In contrast to the views on speed and quality, opinions on the impact on team collaboration were divided. While 53\% see an improvement, over 41\% do not perceive any real change or impact on the team. There were also some negative responses, indicating that team collaboration can deteriorate when using GenAI.




\subsubsection{Evolution of Perceptions: Comparing Initial and Sustained Use}
\label{sec:perception_evolution}




Although the anonymous survey design prevented formal paired analysis of S1--S2 changes at the individual level, comparing aggregate patterns between initial use (S1, $n=27$) and sustained use (S2, $n=17$) characterizes how perceptions evolved across the sample. 
We examine three dimensions measured in both surveys: impact on development speed, software quality, and team collaboration.

\paragraph{Impact on Development Speed}
Perceptions of GenAI's positive impact on development speed remained consistently high across both time points. 
At S1, all 22 participants who had used GenAI agreed that it could accelerate the development process. 
At S2, sustained experience reinforced these initial expectations: 59\% reported GenAI ``significantly improves'' or provides ``transformative'' improvement to development speed, with an additional 35\% reporting ``slight improvement'' and only one participant (6\%) perceiving no change. 
No participants perceived a decline in development speed. 
This stability suggests that initial positive expectations about productivity gains were confirmed through sustained use.

\paragraph{Impact on Software Quality}
Perceptions of quality impact showed more nuance at S2 compared to S1. 
At S1, 68\% who had used GenAI believed GenAI could influence better development practices, while 32\% did not. 
At S2, after six months, this pattern had evolved: 71\% perceived at least slight improvement in software quality, but 18\% now perceived that quality ``slightly worsens'' with GenAI use, and 12\% saw no real change. 
This shift towards more mixed perceptions may reflect increased realism as developers encountered GenAI's limitations in practice (e.g., hallucinations, context-inappropriate suggestions, and the need for careful code review), tempering initial optimism with grounded experience.

\paragraph{Impact on Team Collaboration}
Collaboration perceptions showed the most variability and some deterioration. The S1 responses contained differing opinions with some anticipating improved team discussions and knowledge sharing, while others shared concerns about reduced peer interaction. 
By S2, these concerns appeared partially validated: while 53\% perceived improvement in collaboration, 41\% reported no real change, and one participant (6\%) perceived a decline. 
This pattern aligns with qualitative findings (presented later) that GenAI is used primarily for individual tasks rather than collaborative activities. 
As shown in Fig.~\ref{fig:tasks}, GenAI was rarely used in collaborative contexts such as stand-up meetings (88\% never use) or general team collaboration (71\% never use).

These longitudinal patterns suggest that whilst performance expectations were largely confirmed during sustained use, perceptions of quality and collaboration became more nuanced as developers gained practical experience. 
The high frequency of sustained use---with 41\% using GenAI daily and an additional 47\% using it multiple times per week---indicates that, despite these evolving perceptions, developers continued to find value in GenAI for their individual work activities.

\subsection{Developer Experiences From Interviews and Observations}\label{sec:res-phase1-qualitative}
This section presents findings from interviews with developers and from fieldwork observations (as described in Sect. \ref{interviews} and  Sect.~\ref{ethnographic_fieldwork}). These interviews and observations occurred between the two surveys, thus providing insights into the use of GenAI tools after a period of sustained use. We discuss the tasks for which GenAI is used, including its frequency of use, before discussing the overall experience with GenAI, highlighting the benefits, challenges, and concerns. Perceptions of the quality of GenAI responses are also noted. The section concludes with a discussion of the learnings developers have gained over time regarding effective GenAI usage. 

\subsubsection{What is Generative AI Used For?}
Notably, many of the development tasks shown in Figure~\ref{fig:tasks} were also mentioned by developers in the interviews. Moreover, when discussing the tasks, we start to see how using GenAI saves developers time.

One of the main use cases for GenAI is \textbf{support for repetitive and routine coding} tasks. For instance, P1 mentions automating monotonous data entry: \say{\textit{Generate 100 lines of this thing here}}, indicating tasks previously done manually or through extensive online searching. Similarly, P12 refers to the rapid creation of complex JSON structures, previously considered a \say{\textit{very tedious task}} which is now completed \say{\textit{in one minute instead of seven hours}} (P12).

\textbf{Refactoring and improving} existing code is another common scenario. Although P2 notes they have to explicitly prompt to refactor the code \say{\textit{without altering the logic!}} to limit the risk of error. Additionally, \textbf{debugging} is frequently simplified by GenAI. P11 describes copying terminal errors directly into the chat, facilitating quick identification and resolution of problems.

Other examples provided by developers included: creating mock data and performing data transformations (P5), explaining unfamiliar code or syntax (P26), 
drafting user stories and acceptance criteria (P12); translating files to support multi-lingual websites (P13); and translating Italian into English as required for pull requests (P12).

Notably, it was observed that \textbf{GenAI tools are not used collaboratively}; instead, interactions are usually handled by a single person. In particular, they were not observed as being used during team meetings. 

Developers are more likely to use GenAI tools \textbf{for more straightforward and repetitive} tasks such as \say{\textit{quick scripts}}, highlighting a \textbf{preference for leveraging these tools to accelerate routine tasks rather than working on more complex tasks} as exemplified by P16: \say{\textit{If instead I write something a bit more structured, cleaner, I prefer to use it less, maybe only for confirmation, just to get feedback that says, \say{this thing is fine}}} (P16). The fact that task characteristics influence usage frequency is also reflected in another participant who mentioned, \say{\textit{It depends a lot on the project I’m in and how confident I am, on the technologies I’m using.}} (P13), implying that GenAI tools can be valuable when dealing with new or unfamiliar technologies or tasks.




Apart from using GenAI tools for coding and other software development tasks, participants also frequently use them to \textbf{support non-development tasks} such as writing and formatting emails or messages, creating presentations, or drafting meeting minutes (P2, P26, P1, P16, P11). The tools are also seen as beneficial for language correction and formalization, especially in English communication with clients. A primary motivation, just as for the development-related tasks, is to be more efficient.





\subsubsection{Usage Frequency}
Interview responses on the \textbf{frequency of use varied significantly and were often related to task complexity and task type}. Some participants describe continuous or intensive daily use, emphasizing the integration into their daily workflows; for example, P11 stated: \say{\textit{ChatGPT I use it a lot a lot for work}}, while P1 noted, \say{\textit{whenever I have a doubt, I use it [...] three, four times a day [...] actually, I always have it open}}, indicating constant accessibility and availability as a supporting factor for frequent interaction. Indeed, some developers (P25, P13) expressed interest in incorporating GenAI further into their daily workflows. P25 shared being conscious about all the positive effects and use cases, and \say{\textit{I would like to integrate it more into my workflows}}. However, not all users reported intensive use; some reported less frequent interactions, ranging from \say{\textit{less than once a day}} (P25) to \say{\textit{one or two times a week}} (P20). This varied usage suggests a spectrum of usage linked closely to specific individual and contextual factors.

\subsubsection{Overall Experience}
This section presents findings on professionals' overall experience with GenAI. It serves two purposes: (1) to reinforce the evidence about the perceived benefits of using GenAI, the challenges posed by tool limitations, and concerns about using GenAI at work as reported in previous work~\citep{davila_industry_2024, vaz_pereira_exploring_2025}; (2) to show which UTAUT2 construct they are related to.

\begin{table}[t!]
    \small
    \centering
    \caption{Benefits of Using GenAI. The numbers next to each benefit indicate the number of participants whose interview data supported that benefit.  Rel. refers to the most relevant UTAUT2 construct.}
    \begin{tabular}    {p{0.16\textwidth}p{0.35\textwidth}p{0.35\textwidth}p{0.05\textwidth}}
    \hline
        \textbf{Benefit} & \textbf{Description} & \textbf{Indicative Example} & \textbf{Rel.}\\
        \hline
        Saves time (8) & Tasks can be completed more quickly with the aid of GenAI & Refactoring would take 2 to 3 times longer without ChatGPT (P2) & \PE\\
        Advantageous compared to existing tools (5) & Better than existing developer tools & e.g., In comparison to Google, ChatGPT provides the answer right away (P2), answers are more contextually adjusted (P25) & \PE\\
        Positive impact on workflows (10) & Overall utility of GenAI tools is high and good integration with workflows & Provides a solution path (P19) & \PE\\
         \hline
    \end{tabular}  \label{tab:benefits_genAI_use}
\end{table}

The developers provided numerous benefits (Table~\ref{tab:benefits_genAI_use}) of using GenAI. The \textbf{benefits included time-saving}, as it was quicker to complete tasks with GenAI; helping acquire knowledge, especially in comparison to other tools; helping when stuck; and providing a sense of comfort in an approach, as it can act as a second opinion on a defined approach. These benefits all contribute to improved \pe~as developers can use GenAI to enhance their task performance.

\begin{table}[t!]
    \small
    \centering
    \caption{Challenges Experienced when using GenAI. The numbers next to each challenge or limitation indicate the number of participants whose interview data supported it. Rel. refers to the most relevant UTAUT2 construct.} 
    \begin{tabular}     {p{0.16\textwidth}p{0.35\textwidth}p{0.35\textwidth}p{0.05\textwidth}}
    \hline
        \textbf{Challenge} & \textbf{Description} & \textbf{Indicative Example} & \textbf{Rel.}\\
        \hline
        Complex and context specific tasks (9) & GenAI lack of awareness of the context leads to suboptimal results and frustration. & GenAI is not optimal for business logic (P20) & \EE\\
        Inaccurate responses (4) & Provides inaccurate answers or responding even when GenAI is unsure & P26 finds GenAI wastes their time with incorrect information & \EE\\
        Unhelpful for specialized technologies (3) & GenAI does not work well for specialized libraries or proprietary languages  & Cannot use for a client's proprietary language (P5) & \PE\\        
        Verbose responses (2) & Responses are overly long, detailed, and contain redundant information  & GenAI repeatedly states basic steps such as install the library (P5) & \EE\\
        Lack of alternative answers (1) & Only presents one solution rather than multiple alternatives to allow the developer to choose the most preferred.  & P8 believes GenAI should offer multiple solutions by default and explain why a specific approach is preferred  & \EE\\
         \hline
    \end{tabular}   \label{tab:challenges_genAI_use}
\end{table}

However, in using GenAI, developers faced challenges (Table~\ref{tab:challenges_genAI_use}) that made it difficult for GenAI to provide satisfactory responses, thus negatively influencing \ee. Challenges that required effort to address (e.g., by writing multiple prompts) include \textbf{ChatGPT missing the broader task context} and challenges with the output including inaccurate and overly responses. Moreover, GenAI does not work well with specialized technologies that some developers are required to use for client projects, leading to a negative impact on \pe. In an extreme case, one participant could not use GenAI due to its lack of knowledge of the specialized language (P5). Finally, one concern was that GenAI tools offer only a single alternative, potentially leading to suboptimal solutions. Workarounds to address this included authoring multiple prompts or resetting the chat.

\begin{table}[t!]
    \small
    \centering
    \caption{Concerns about using GenAI. The numbers next to each concern indicate the number of participants whose interview data supported that concern. Rel. refers to the most relevant UTAUT2 construct.}
    \begin{tabular}     {p{0.16\textwidth}p{0.35\textwidth}p{0.35\textwidth}p{0.05\textwidth}}
    \hline
        \textbf{Concerns} & \textbf{Description} & \textbf{Indicative Example} & \textbf{Rel.}\\
        \hline
        Unreliable results (11) & Expressing scepticism on the reliability of results leading to a lack of trust in GenAI's responses & Many developers verify and double-check the AI responses as acutely aware \say{\textit{the hallucination is always round the corner}}(P19) & \EE \\
        Over-reliance (9) & Concern that over-reliance on GenAI can hinder creativity, learning, and personal growth & Concerns it becomes a substitute for personal learning (P16), novices unable to identify \say{\textit{code that is okay but not optimal}} (P2) & \EE\\
        Code quality and insecure code (3) & GenAI can generate lower quality code and/or more insecure code compared to manually written code & Tool generates generic solutions, leading to a \say{\textit{flattening}} (P2) of the code base, developer under pressure may overlook security vulnerabilities introduced by AI (P16) & \PE\\
        Privacy and confidentiality risks (6) & Concerns that sensitive client data may be inadvertently shared and used for model training& Discomfort sharing code containing client information (P1) & \EE\\
         \hline
    \end{tabular}  
    \vspace{-2mm}
    \label{tab:concerns_genAI_use}
\end{table}

Developers also raised several concerns (Table~\ref{tab:concerns_genAI_use}) about using GenAI in their workplace. A \textbf{significant concern was unreliable responses} leading to a lack of trust in GenAI tools, along with a concern that over-reliance can degrade one's skill set. Concerns about poor code quality and insecure code were also raised, along with potential privacy and confidentiality risks when using GenAI in clients' codebases. In the case of knowledge-seeking, some developers (e.g., P8, P11, P20, P25) preferred established resources (Google, Stack Overflow, official documentation) because they found it challenging to shift their habitual mindset, especially as concerns about unreliable or unsatisfactory GenAI responses lingered. Effectively, GenAI served as a complement to these traditional platforms. These concerns can negatively affect the \ee and \pe~when using GenAI.

Many of these benefits, challenges, and concerns align with those reported in~\cite{davila_industry_2024} and \cite{vaz_pereira_exploring_2025}. These two studies also found that a benefit of using GenAI was that it helped complete tasks more quickly, while noting challenges around poor performance on complex tasks, concerns about the privacy and confidentiality of data, the reliability of responses, and the risk of over-reliance. This similarity across the studies highlights that, with the adoption and sustained usage of GenAI, the developer experience is changing in similar ways, irrespective of context.

\subsubsection{Perceptions about the Quality of GenAI responses}
In discussing their use of GenAI tools, it was noticeable that participants held \textbf{varying views on the quality of the responses} from the tools. Some found the quality good, with P10 noting they were \say{\textit{Quite a lot, very.}} satisfied with the quality of the responses. Others disagreed, considering the responses poor (see the challenges noted in Table~\ref{tab:challenges_genAI_use}). Some noted the quality of the response varied depending on the task for which assistance was sought. For example, P1 felt \say{\textit{[On a scale] from 1 to 10, a 7, a 7.5. Maybe there’s a task where it gives you a 9 answer, and another task where it really goes off-track and maybe it’s only a 5.}}. This variance perhaps points to a recognition that GenAI is not suitable for all tasks and developers are still learning where it excels.

\subsubsection{Learning to use GenAI effectively}
As the developers had been using GenAI tools for some time, they described several key learnings that helped them use them more effectively. This learning helped to overcome some of the challenges and limitations noted in Table~\ref{tab:challenges_genAI_use}. Some of these learning relate to the roles of social influence and facilitating conditions that help developers make better use of GenAI.

\textbf{Prompt crafting} emerged as a critical factor influencing GenAI effectiveness. Developers enhanced their prompting skills through trial and error, relying on experiential rather than formal learning (P16). They emphasized structured prompts, explicit contextualization, and iterative refinement (P2, P12, P16, P19). For example, specific keywords such as \say{\textit{You are this}}, or \say{\textit{Give me a step-by-step solution}}, significantly improve outcomes (P2). 

Another learning was \textbf{identifying the tasks that GenAI tools perform effectively}. While GenAI tools effectively handle clearly defined tasks, they struggle with complex business logic. P20 stated, \say{\textit{if I need to implement [...] business logic, no!}}. However, this level of awareness regarding the practical uses and capabilities of GenAI varied. While some developers clearly recognized multiple opportunities to leverage these tools, others seemed less aware of potential applications and use cases, and consequently used the tools for only a limited set of use cases. 




The developers \textbf{learned to balance the time invested in careful prompt writing against the potential time savings} from AI assistance. The developers acknowledged that crafting precise prompts significantly improves the quality of GenAI outputs. P2 noted: \say{\textit{I realize that to get good results, it's necessary to write a good prompt,}} yet added, \say{\textit{sometimes I choose in the tradeoff of the time needed to write a good prompt, to not waste that much time.}} 
Poorly written prompts frequently led to unsatisfactory responses, requiring additional clarifications or repeated queries. 

Learning from others by \textbf{knowledge exchange varied significantly} across developers and teams. Several participants reported informal discussions with colleagues about best practices and tips for using these tools. For example, P26 mentioned that if \say{\textit{one finds an AI tool that he likes that he finds useful he shares it with others}}, thus facilitating collective learning. In contrast, P1 noted that such exchanges are less frequent for them because they do not come to the office often. This implies that such exchanges occur more often when team members meet in person; consequently, developers who work primarily from home might have fewer opportunities for knowledge exchange. 

However, the \textbf{practice of directly teaching others is limited}. For P1, such instances were rare. 
This contrasts with more active teams, such as P2 and P13, which reported regular discussions within their teams about how to use GenAI tools effectively, leading them to try a GenAI tool to translate a file. 


Beyond social interactions with colleagues, another way to gain knowledge about effective use is through training. However, there were \textbf{mixed views on the need for training} in GenAI, with some advocating training (P1, P5, P12, P25) and others noting that developers should educate themselves or learn on the job (e.g., P5). 

In summary, it appears that a combination of colleagues' social influence and facilitating conditions, such as time for training and experiential learning on the job, helps developers increase their self-efficacy in using GenAI tools.

\subsection{UTAUT2 Constructs Assessment from the Sustained Use Survey}\label{sec:res-phase1-quantitative}

Having established baseline usage patterns and perceptions in Sect.~\ref{sec:res-phase1-usage}, we now examine how well the UTAUT2 theoretical framework captures the factors influencing sustained GenAI use at Apuliasoft. 
This subsection presents quantitative validation of all seven UTAUT2 constructs through descriptive statistics, bivariate correlations, and examination of construct-specific response patterns, enabling comprehensive exploratory analysis.
These exploratory findings informed both our theoretical model specification and instrument refinements for Phase~2.

\subsubsection{Descriptive Statistics and Response Patterns}

Table~\ref{tab:utaut2_descriptives} presents descriptive statistics for all the UTAUT2 constructs measured in the sustained use survey (S2).
These statistics characterise perceptions after six months of sustained use but, given the small sample size (n=17), should be interpreted as exploratory rather than confirmatory evidence.

\begin{table}[tb]
\small
\centering
\caption{Descriptive statistics of UTAUT2 constructs from sustained use survey S2 (n=17).}
\begin{tabular}{lccc}
\hline
\textbf{Construct} & \textbf{Items} & \textbf{Mean} & \textbf{SD} \\
\hline
Performance Expectancy (PE) & 4 & 4.09 & 0.45 \\
Effort Expectancy (EE) & 4 & 4.07 & 0.41 \\
Social Influence (SI) & 3 & 3.49 & 0.75 \\
Facilitating Conditions (FC) & 4 & 4.12 & 0.44 \\
Hedonic Motivation (HM) & 3 & 3.57 & 0.70 \\
Price Value (PV) & 3 & 3.00 & 0.93 \\
Habit (H) & 3 & 2.98 & 0.95 \\
Continued Use Intention (CUI) & 3 & 4.02 & 0.61 \\
\hline
\multicolumn{4}{l}{\footnotesize Scale: 1 = Strongly Disagree to 5 = Strongly Agree}
\end{tabular}
\label{tab:utaut2_descriptives}
\end{table}

Three constructs showed particularly high means, namely \pe (M = 4.09, SD = 0.45), \ee (M = 4.07, SD = 0.41), and \fc (M = 4.12, SD = 0.44). 
These high scores suggest that after six months of sustained use, developers strongly perceived GenAI as useful, easy to use, and well-supported by available resources and knowledge. 
The relatively low standard deviations indicate reasonable consensus among participants, though this interpretation remains tentative given the sample size.

\si is lower (M = 3.49, SD = 0.75), suggesting that external social pressures played a relatively modest role in sustained use decisions. 
This finding aligns with Apuliasoft's organisational context, where GenAI adoption occurred organically without formal mandates or structured peer pressure. 
\hm fell in the moderate range (M = 3.57, SD = 0.70), indicating that whilst developers found GenAI enjoyable, intrinsic enjoyment was not the primary driver of continued use.

\pv (M = 3.00, SD = 0.93) and \hab (M = 2.98, SD = 0.95) showed the lowest means of all constructs.
The low \hab mean reflects that GenAI use remained a deliberate, context-dependent choice rather than an automatic behaviour after six months of exposure. 
The modest \PV scores confirm that this construct is less salient in the organizational context. 
Notably, these two constructs also showed higher standard deviations, indicating more dispersed perceptions compared to the other UTAUT2 dimensions.

Finally, \cui was high (M = 4.02, SD = 0.61), consistent with observed usage patterns and workflow integration documented in subsection 5.2.

\begin{figure}
    \centering
    \includegraphics[width=1\linewidth]{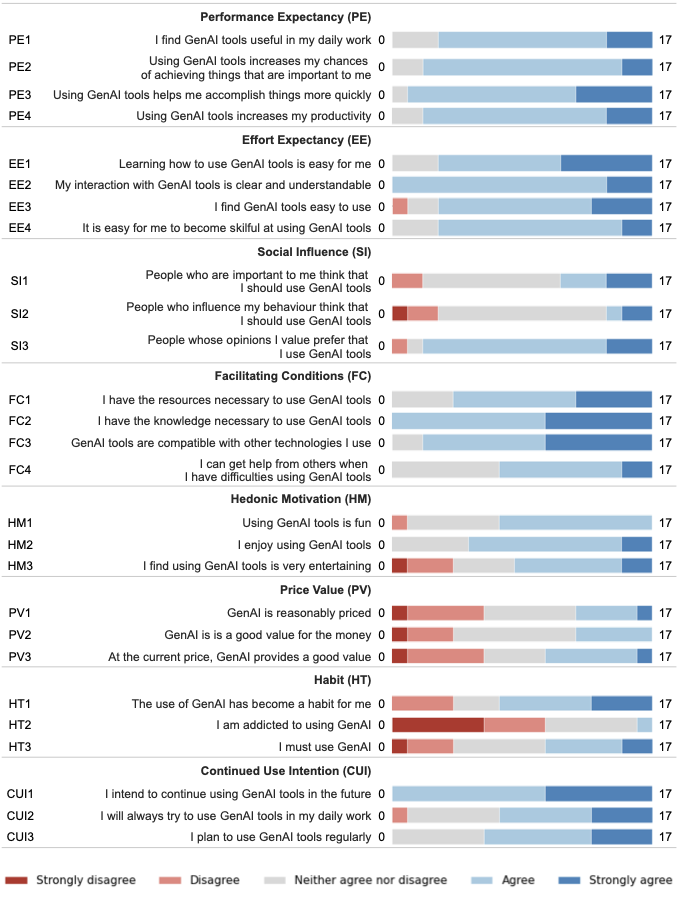}
    \caption{Response distributions for each UTAUT2 construct from the survey S2 (n=17).}
    \label{fig:s2_utaut2_distribution}
\end{figure}

To understand these aggregate statistics more deeply, we examined response distributions for each construct, illustrated in Figure~\ref{fig:s2_utaut2_distribution}. The \pe received strong endorsement across all items, with developers expressing clear agreement that GenAI tools improve productivity, enhance work performance, and accelerate task completion. 
This construct emerged as one of the strongest predictors of continued use, consistent with UTAUT2 predictions and complementary empirical findings on GenAI adoption in software development contexts~\citep{vaz_pereira_exploring_2025}.

\ee perceptions were similarly positive (Figure~\ref{fig:s2_utaut2_distribution}), indicating that Apuliasoft developers perceived GenAI tools as relatively easy to learn and use. 
This signals that the cognitive overhead associated with adoption remained low even after six months of sustained use, reinforcing findings from recent studies on developers' quick uptake of GenAI-based assistants. 
The strong agreement across effort-related items suggests that usability concerns did not emerge as barriers during the maturation period.


\fc (Figure~\ref{fig:s2_utaut2_distribution}) showed moderate-to-high agreement, suggesting that developers felt they possessed adequate knowledge, resources, and organisational support to use GenAI tools effectively. 
However, the item regarding asking others for help when experiencing difficulties received approximately 41\% neutral responses. 
This pattern indicates that while the technical infrastructure appeared sufficient, peer-support networks around GenAI use were still developing within the organisation. 
This finding resonates with qualitative data presented earlier in Sect.\ref{sec:res-phase1-qualitative}, where some participants noted limited knowledge sharing about GenAI practices, particularly in remote work settings.


\si (Figure~\ref{fig:s2_utaut2_distribution}) showed a predominantly neutral pattern, with gradually emerging positive social encouragement surrounding GenAI tools. 
The strongest endorsement appeared for the item about people whose opinions participants value recommending tool use, with $\sim 88\%$ agreement. 
This suggests that trusted peers played an influential role in reinforcing tool adoption, even in the absence of formal organizational pressure. 
The remaining items exhibited more dispersed responses, with neutral selections as the most frequent answer, indicating that broader social norms around GenAI use were still forming within the organization. 
This finding aligns with the voluntary adoption context at Apuliasoft, where developers independently chose whether and how to integrate GenAI into their workflows.

\hm (Figure~\ref{fig:s2_utaut2_distribution}) revealed a nuanced pattern. 
Respondents consistently reported enjoyment when interacting with GenAI tools, though they did not necessarily view them as entertaining. 
No participants expressed negative perceptions regarding the statement about liking the tool, and only one negative response appeared for the item about using the tool being fun. 
These results suggest that developers derived intrinsic satisfaction from using GenAI tools, but this satisfaction was rooted in productive engagement rather than leisure or entertainment value.
This distinction proved important for survey instrument refinement, as discussed later in Sect.~\ref{sec:res-phase1-survey-refinements}.


Regarding \hab, results indicate that habitual use of GenAI tools is emergent but not yet consolidated. While 35.3\% of respondents agreed and 23.5\% strongly agreed that using the tool has become a habit (total 58.8\% positive), a substantial proportion remained neutral (17.6\%) or disagreed (23.5\%). Similarly, 41.2\% reported that they tend to use the tool automatically, but 35.3\% remained neutral, suggesting that routine utilization is still varying across developers. Notably, 58.8\% disagreed with the statement “I feel strange if I don’t use the tool”, indicating that GenAI use has not yet become indispensable to daily work.

Responses on \pv (Figure~\ref{fig:s2_utaut2_distribution}) reveal a mixed evaluation. For the item “The cost of using the tool is acceptable”, 35.3\% agreed and 11.8\% strongly agreed, while 29.4\% disagreed. A similar pattern appears for perceived price reasonableness, with 23.5\% agreement and 35.3\% neutrality. Regarding value for money, nearly half of respondents (47.1\%) are neutral, indicating uncertainty rather than outright rejection.



Overall, the seven constructs demonstrated that UTAUT2 provides a suitable explanatory framework for understanding developers' sustained use perceptions in this setting. 
The results confirmed the relevance of the theoretical model while also revealing context-specific patterns that informed Phase~2 instrument refinements and hypothesis priorities.

\subsubsection{Bivariate Correlations Between UTAUT2 Constructs}
\label{sec:res-phase1-correlations}

To explore relationships between UTAUT2 constructs and Continued Use Intention, we calculated Pearson correlations using the responses to the sustained use survey S2 (see Table~\ref{tab:correlation_matrix} in Appendix ). 
Given the small sample size (n=17), these correlations should also be interpreted as exploratory patterns that informed our expectations for the confirmatory Phase~2 study rather than as definitive evidence.

\pe showed the highest correlation with \cui ($r = 0.30$), though this did not reach statistical significance ($p = .24$) given the limited statistical power. 
Nevertheless, the direction remains consistent with UTAUT2 theory, which positions \PE as typically the strongest predictor of behavioral intention.

\fc ($r = 0.15$) and \si ($r = 0.13$) showed weak positive associations with \CUI. 
Notably, \ee ($r = -0.03$), \hm ($r = -0.08$), \pv ($r = 0.05$), and \hab ($r = 0.17$) all exhibited weak or near-zero correlations with the outcome variable. 
The weak relationships for \EE and \HM may reflect that ease of use and enjoyment become less salient for sustained use once initial adoption hurdles have been overcome. 
The weak correlations for \PV and \HAB provided empirical support for their exclusion from the Phase~2 theoretical model, complementing the theoretical and qualitative rationale detailed in Sect.~\ref{sec:phase1-constructs-exclusion}.

One significant correlation emerged between \EE and \SI ($r = 0.50$, $p < .05$). 
This relationship suggests that developers who found GenAI easy to use also perceived greater social encouragement.
One plausible explanation is that ease of use facilitates peer discussions and recommendations, as developers who successfully integrated GenAI into their workflows were more likely to share experiences and encourage colleagues.

These exploratory correlation directions aligned broadly with existing literature patterns and informed our hypothesis prioritisation for Phase~2. 
The patterns suggested that Performance Expectancy would likely emerge as a key predictor, whilst other constructs might show weaker direct effects on \cui in the voluntary SME context we studied.

\subsection{Triangulating Evidence Across Data Sources}

This section triangulates the S1 and S2 survey data with the interview and observation data to discuss the Apuliasoft case study evidence supporting the relevance of each UTAUT2 construct to sustained use. Multiple data sources indicate that \pe and \ee may be highly relevant to \cui. Other constructs are also relevant, but perhaps less so than these.

\paragraph{\pe~(\PE)}
Both surveys find that nearly all (except one) respondents who use GenAI tools perceive that GenAI tools speed up development. Additionally, the second survey indicates that many developers believe that GenAI improves code quality. Moreover, when considering performance expectations directly, 89\% believe that GenAI improves their work performance, and the same percentage believe it increases their productivity. The interviews and observations also support this perception, with one notable benefit being that development speeds up. There is therefore strong support for the hypothesis (H1) that a high \pe will lead to \cui GenAI. 

\paragraph{\ee~(\EE)}
The two surveys do not directly assess effort expectancy beyond the questions in S2 that relate to the UTAUT2 effort expectancy construct. As can be seen in the descriptive statistics (Table~\ref{tab:utaut2_descriptives}), developers confirm that the required effort is low. While the interviews did not directly address effort expectancy, it is notable that developers use GenAI from daily to less frequent use. This perhaps indicates that GenAI has been integrated into the workflow for those who use it daily, and so the effort expectancy is low. For others, it is unclear if the effort required is too high leading to low usage, or as noted in challenges, for other reasons, such that the nature of their work makes GenAI unsuitable (e.g., they work on proprietary technologies). Together, the survey and interview data support the hypothesis (H2) that if GenAI is easy to use (low effort expectancy), then it is used more frequently.

\paragraph{\si~(\SI)}
In contrast to the prior two constructs, survey and interview data indicate that social influence does not have a strong effect on sustained use. The results of the UTAUT2 construct in the second survey show that the social influence is low (Tables~\ref{tab:utaut2_descriptives} and  ~\ref{tab:correlation_matrix}). This low level of social influence is somewhat evident in the interviews and observations, if we consider sharing knowledge as a form of social influence; for example, discussing how well GenAI performs on certain tasks can influence other developers to adopt it. The level of knowledge sharing about GenAI best practices was mixed, with some teams sharing extensively and others less so. It therefore appears that \si is less critical for sustained GenAI use compared to the other constructs.

\paragraph{\hm~(\HM)}
S2, in its UTAUT2 questions, was the only instrument that explicitly captured participants' perceptions of the importance of hedonic motivation in using GenAI. As seen in (Tables~\ref{tab:utaut2_descriptives} and  ~\ref{tab:correlation_matrix}), there is limited support for the importance of hedonic motivation in sustained use of GenAI, especially compared to other factors such as \pe and \ee. 

\paragraph{\fc~(\FC)}
Both S2 and the surveys and observations examined facilitating conditions. S2, in its UTAUT2 construct, found a weak positive correlation between facilitating conditions and sustained use (Table~\ref{tab:correlation_matrix}). This weak support is also evident in the interviews, which revealed contrasting views on the importance of training (as part of creating an environment that facilitates use) to help developers make more effective use of GenAI. Some participants wanted formal training, whereas others expected to learn on the job or by sharing knowledge with colleagues. These varying perspectives perhaps highlight that facilitating conditions differ from individual to individual, yet they remain an essential consideration. There is thus some evidence supporting the influence of \fc on \cui.

\subsection{Justification for Excluded Constructs}\label{sec:phase1-constructs-exclusion}

Although the sustained use survey (S2) measured all seven UTAUT2 constructs to enable a comprehensive assessment, convergent evidence supported excluding \hab and \pv from the confirmatory model tested in Phase~2.

The IS Continuance Model~\citep{limayem2007habit} positions \hab (\HAB) as a potential outcome of sustained use rather than a predictor of continued use intention.
Consistent with this theoretical framing, developers described their GenAI use as deliberate and context-\hspace{0pt}dependent rather than habitual.
Usage varied based on task type (``\textit{It depends a lot on the project I'm in}''; P13), technology familiarity, and time pressure. 
\hab items showed the lowest mean (2.98) of all constructs, with 47\% disagreeing with the item ``\textit{I feel strange if I don't use the tool}.'' 
This pattern reflects the voluntary, emergent nature of GenAI adoption at Apuliasoft, where, without organisational mandates, use remained a conscious choice rather than an ingrained habit.

\pv (\PV) captures consumers' cognitive trade-off between perceived benefits and monetary cost~\citep{venkatesh2012utaut2} as it was developed for consumer contexts where individuals directly bear financial costs.
\PV considerations were largely absent from the interview discussions, and when the topic arose, participants
did not discuss cost-benefit trade-offs.
In the enterprise context, where developers do not bear subscription costs, \PV lacks relevance as a predictor of continued use.

This convergent pilot evidence justified excluding \hab and \pv from the theoretical model, focusing instead on the five constructs that demonstrated relevance in the voluntary SME context.

\subsection{Survey Instrument Refinements}
\label{sec:res-phase1-survey-refinements}

The pilot case study informed several refinements to the survey instrument deployed in Phase~2. 
Specifically, the analysis of sustained use survey response distributions, combined with qualitative feedback from interviews, prompted specific item adjustments to improve clarity and contextual appropriateness.

Interview data revealed that developers used varied terminology when discussing GenAI tools (e.g., ChatGPT, Copilot, AI assistants), leading us to provide concrete examples alongside generic references to ``GenAI tools'' in survey items to ensure consistent interpretation across respondents.
The \hm item {\HM}\texttt{3} regarding whether ``the tool offers entertainment'' showed the most dispersed responses, with many participants selecting neutral options. 
Qualitative data clarified that developers found GenAI \textit{enjoyable} in a professional sense but not entertaining in the way the term might imply leisure or gaming contexts.
For Phase~2, this item was reworded to better capture productive enjoyment rather than `entertainment' value, ensuring the construct measured intrinsic satisfaction from work-related tool use rather than recreational appeal.
The \si item regarding people who influence behaviour received predominantly neutral responses ({\SI}\texttt{2}), possibly because developers perceived their GenAI use as a personal professional choice rather than one influenced by authority figures. 
This item was retained in Phase~2 but rephrased to provide additional context by explicitly listing examples of such influencers (e.g., team leads, managers, mentors). 
This adjustment aimed to help respondents recognise relevant social influences that might not have been immediately apparent in the original phrasing.

Beyond these specific item modifications, the core survey structure was carried forward to Phase~2 with minimal changes. 
The anonymous survey design was maintained across both phases. 
Similarly, the five-point Likert scale format for all UTAUT2 items was retained to enable direct comparison between the two study phases while following established UTAUT2 methodology. 
These procedural consistencies allowed Phase~2 to build directly upon Phase~1 findings while testing the theoretical model with adequate statistical power across multiple Italian SMEs.

\subsection{Summary of Phase 1 Findings}


\paragraph{GenAI is well-integrated into SME developer workflows} After six months of sustained use, 88\% of developers used GenAI daily or multiple times a week, with 89\% reporting good workflow integration.
This confirms our study captures continued use dynamics rather than initial adoption.

\vspace{-6px}
\paragraph{Performance expectations drive sustained use} 
Quantitative data showed PE with the highest mean ($4.09$) and the strongest correlation with \CUI ($r = 0.30$).
In line with technology adoption research, qualitative evidence overwhelmingly emphasised productivity and time-saving benefits.

\vspace{-6px}
\paragraph{Longitudinal patterns suggest perception maturation} 
Cross-sectional comparison of S1 and S2 suggests that initial positive expectations about development speed were confirmed, while perceptions of quality and collaboration became more nuanced with sustained experience. These nuances were supported by the interviews and observations, in which developers indicated they had learned to use GenAI effectively through sustained use. 

\vspace{-6px}
\paragraph{The five-construct UTAUT2 model is appropriate for this context} 
All five included constructs (\PE, \EE, \SI, \HM, \FC) showed theoretical relevance supported by survey and interview data. 
The exclusion of \HAB and \PV was validated by low pilot scores and the absence of qualitative themes.

\vspace{-6px}
\paragraph{Survey instrument refinements informed Phase~2 deployment} 
Minor refinements to item wording and translation improved clarity without substantive changes to construct operationalisation. 
The core measurement structure, using five-point Likert scales and an anonymous design, was retained for Phase~2 to enable direct comparison with pilot findings.

\vspace{4px}
The findings of the exploratory pilot study at Apuliasoft informed the design of the validation Phase~2 study, 
which tests the hypothesised structural relationships across a larger sample of SMEs.

\section{Results of Phase 2: Cross-sectional Validation Study} \label{sec:phase2-results}

\subsection{Sample Characteristics}
\label{sec:sample-characteristics}

The final analytical sample comprised 154 valid responses from software developers who actively use GenAI tools in their professional work (see Sect.~\ref{sec:phase2-data-collection}).
Table~\ref{tab:S3-demographics} summarises the sample demographics.
The respondents were predominantly male (92.9\%).
The majority of participants were aged 26--45 (61.0\%), with 33.1\% aged 26-35, 27.9\% aged 36-45, and 7.1\% aged 56 or older.
The sample was professionally experienced: 76.0\% had four or more years of software development experience, and over half (51.9\%) had eight or more years.

To verify that our sample represents voluntary rather than mandatory adoption contexts, we measured voluntariness of use through a four-item scale based on~\citet{moore1991isr-development} and incorporated in the UTAUT framework.
Among respondents with valid data on these items ($n = 119$), about 94\% agreed or strongly agreed that their decision to use GenAI was entirely voluntary, and 79\% agreed that their employer does not require them to use such tools.

Regarding GenAI usage patterns, the sample represented active adopters with substantial tool experience.
Over 66\% had been using GenAI tools for at least six months, and about 31\% had used them for more than one year.
Usage frequency was high: 63.6\% used GenAI tools at least several times per week, with 31.2\% reporting daily use.
This usage profile confirms that participants had moved beyond initial experimentation and were suitable for studying continued use intentions.

\begin{table}[tb]
\small
\centering
\caption{Phase~2 (S3) survey sample demographics (n = 154)}
\label{tab:S3-demographics}
\begin{tabular}{llrr}
\hline
Category & Value & n & \% \\
\hline
Professional Experience & Less than 1 year & 8 & 5.2 \\
 & 1--3 years & 29 & 18.8 \\
 & 4--7 years & 37 & 24.0 \\
 & 8+ years & 80 & 51.9 \\
\hline
Age & 18--25 & 29 & 18.8 \\
 & 26--35 & 51 & 33.1 \\
 & 36--45 & 43 & 27.9 \\
 & 46--55 & 20 & 13.0 \\
 & 56+ & 11 & 7.1 \\
\hline
Gender & Male & 143 & 92.9 \\
 & Female & 11 & 7.1 \\
\hline
GenAI Usage Duration & Less than 1 month & 10 & 6.5 \\
 & 1--6 months & 43 & 27.9 \\
 & 6--12 months & 53 & 34.4 \\
 & More than 1 year & 48 & 31.2 \\
\hline
GenAI Usage Frequency & Rarely (monthly or less) & 12 & 7.8 \\
 & Occasionally (several times/month) & 44 & 28.6 \\
 & Frequently (several times/week) & 50 & 32.5 \\
 & Daily & 48 & 31.2 \\
\hline
\end{tabular}
\end{table}

Table~\ref{tab:descriptives} presents the descriptive statistics for the model constructs.
All constructs exhibited means above the scale midpoint (3.0), indicating generally favourable perceptions of GenAI tools.
\cui showed the highest mean (M = 4.16, SD = 0.61), followed by \pe (M = 4.10, SD = 0.68), suggesting that developers perceive substantial performance benefits and intend to continue using GenAI tools.
\fc (M = 4.02, SD = 0.64) and \ee (M = 4.01, SD = 0.69) also scored highly, indicating that developers find these tools easy to use and perceive adequate organisational support.
\si showed the lowest mean (M = 3.50, SD = 0.53), suggesting moderate rather than strong social pressures to use GenAI, consistent with the voluntary adoption context.
\hm (M = 3.76, SD = 0.71) indicated moderate enjoyment from using GenAI tools.
All constructs demonstrated acceptable variance, with standard deviations between 0.53 and 0.71, confirming sufficient response variability for structural equation modelling.
Skewness values ranged from $-0.05$ to 0.64 and, although PLS-SEM is robust to non-normality~\citep{hair2019plssem}, these values indicate near-normal distributions.

\begin{table}[tb]
\small
\centering
\caption{Descriptive statistics for model constructs (n = 154)}
\label{tab:descriptives}
\begin{tabular}{lcccccc}
\hline
Construct & Items & Mean & SD & Min & Max & Skew \\
\hline
Performance Expectancy & 4 & 4.10 & 0.68 & 3.00 & 5.00 & $-$0.05 \\
Effort Expectancy & 4 & 4.01 & 0.69 & 3.00 & 5.00 & 0.10 \\
Social Influence & 3 & 3.50 & 0.53 & 2.33 & 5.00 & 0.64 \\
Hedonic Motivation & 3 & 3.76 & 0.71 & 1.67 & 5.00 & 0.05 \\
Facilitating Conditions & 4 & 4.02 & 0.64 & 3.00 & 5.00 & 0.11 \\
Continued Use Intention & 3 & 4.16 & 0.61 & 3.00 & 5.00 & 0.00 \\
\hline
\multicolumn{7}{l}{\parbox{0.9\textwidth}{\footnotesize\textit{Note.} All items measured on 5-point Likert scales (1 = Strongly Disagree to 5 = Strongly Agree).}}\\
\vspace{-5mm}
\end{tabular}
\end{table}

\paragraph{Preliminary scale validation}\label{sec:phase2-efa}
Because several UTAUT2 scales were adapted to the GenAI sustained use context, we conducted Exploratory Factor Analysis (EFA) using listwise deletion ($n = 149$ complete cases) to verify the factor structure before proceeding to PLS-SEM.
As detailed below, measurement issues with \SI led to its exclusion from the final structural model, reducing the tested predictors from five to four.

The Kaiser-Meyer-Olkin measure of sampling adequacy was 0.875, indicating factorability, and Bartlett's test of sphericity was statistically significant ($\chi^{2}$ = 2176.42, df = 210, $p < .001$), confirming sufficient shared variance for factor analysis.
Parallel analysis suggested a two-factor solution, which we extracted using oblique (oblimin) rotation, given the expected correlations among UTAUT2 constructs.
The rotated solution explained 51.58\% of total variance, below the conventional 60\% threshold; however, given the strong sampling adequacy and significant correlations, we proceeded with interpretation.

The two-factor EFA solution revealed distinct groupings (see Table~\ref{tab:efa_pattern} in \ref{app:phase2}).
Factor 1 comprised all \PE items, all \HM items, and all \CUI items, with strong primary loadings (0.58--0.85).
Factor 2 comprised all \EE items and all \FC items, again with strong loadings (0.57--0.86).
The \SI items proved problematic: SI1 and SI2 exhibited weak loadings (0.29 and 0.37) with high uniqueness values (0.89 and 0.85), while SI3 showed only marginal loading (0.49) with high uniqueness (0.73).
These results suggest that \SI does not function as a coherent construct in the GenAI sustained use context.
Based on these findings and consistent poor performance of \SI across both the Phase~1 pilot and Phase~2 EFA, we excluded \SI from subsequent PLS-SEM analysis.
\textbf{The final model thus includes the four predictors} \PE, \EE, \HM, and \FC, measured by 17 items.
Additionally, FC4 exhibited high uniqueness (0.67) in the EFA and was removed to improve the measurement model.

\paragraph{Common method bias assessment}
Because all constructs were measured using self-report items from the same survey instrument at the same point in time, we assessed the potential for common method bias (CMB)~\citep{russo_pls-sem_2021}.
We conducted Harman's single-factor test by subjecting all 17 construct items to an unrotated principal component analysis~\citep{podsakoff2003cmb}.
The first factor explained 40.23\% of the total variance, below the 50\% threshold commonly used to indicate substantial CMB.
Furthermore, four factors emerged with eigenvalues exceeding 1.0.
These results suggest that CMB is not a threat to the validity of our findings.
  
\subsection{Measurement Model Assessment}\label{sec:phase2-measurement}

Following established PLS-SEM guidelines~\citep{hair2019plssem, russo_pls-sem_2021}, we assessed the measurement model by examining internal consistency reliability, convergent validity, and discriminant validity for all reflective constructs.
Detailed results are provided in \ref{app:measurement-model}.

\paragraph{Internal Consistency and Convergent Validity}
Internal consistency---i.e., the degree to which indicators measure the same construct---was evaluated using Cronbach's alpha ($\alpha$) and composite reliability (CR) (see Table~\ref{tab:measurement_model}).
All constructs exceeded the recommended $\alpha \geq 0.70$ threshold, with values ranging from 0.780 (\HM ) to 0.904 (\PE).
However, Cronbach's alpha tends to underestimate reliability by assuming equal indicator loadings, and composite reliability is the preferred measure in PLS-SEM~\citep{hair2019plssem}.
All constructs demonstrated strong CR values (0.869--0.937).

Convergent validity---i.e., the extent to which a construct explains the variance of its indicators---was assessed through indicator loadings and Average Variance Extracted (AVE).
All AVE values exceeded the 0.50 threshold, ranging from 0.690 (\HM) to 0.832 (\CUI), indicating that each construct explains more than half of its indicators' variance (Table~\ref{tab:measurement_model}).
Indicator loadings represent the correlation between an item and its construct; values $\geq 0.708$ indicate the item shares at least 50\% of its variance with the construct.
The majority of loadings exceeded this threshold, yet two indicators showed marginally lower values (\EE\texttt{3} = 0.77; \HM\texttt{3} = 0.73).
Following \citet{russo_pls-sem_2021}, we retained these items as their loadings exceeded 0.40 and their removal would not substantially improve AVE, while their theoretical contribution warranted retention.

\paragraph{Discriminant Validity}

Discriminant validity was assessed using the Heterotrait-Monotrait ratio (HTMT) and the Fornell-Larcker criterion. 
Overall, the measurement model demonstrates strong validity.

HTMT assesses discriminant validity by comparing correlations between indicators of different constructs to correlations within constructs; values approaching 1 indicate the constructs are not distinct. All HTMT values (see Table~\ref{tab:htmt}) fell below the conservative 0.85 threshold~\citep{henseler2015htmt}, with the highest value observed between \EE and \FC (0.827). This moderate association is theoretically expected, as both constructs relate to the ease and support for technology use, yet they remain distinct.

The Fornell-Larcker criterion tests whether a construct shares more variance with its own indicators than with other constructs, verified by comparing the square root of each construct's AVE against its inter-construct correlations (Table~\ref{tab:fornell_larcker}). All constructs satisfied this criterion, confirming they are distinct.



\subsection{Structural Model Assessment}\label{sec:phase2-structural}

Having established adequate measurement properties, we proceeded to assess the structural model.
Following PLS-SEM guidelines~\citep{hair2019plssem, russo_pls-sem_2021}, we evaluated collinearity among predictors, path coefficient significance via bootstrapping, coefficient of determination ($R^2$), effect sizes ($f^2$), and predictive relevance ($Q^2$).

\paragraph{Collinearity Assessment}

We first examined collinearity among the predictor constructs using Variance Inflation Factors (VIF).
All VIF values were well below the conservative threshold of 5: PE (1.75), EE (2.20), HM (1.46), and FC (2.54).
The highest VIF (2.54 for FC) indicates that collinearity is not a concern for interpreting the structural model results.

\paragraph{Path Coefficients and Hypothesis Testing}

We assessed path coefficient significance using bootstrapping with 5,000 subsamples, following recommended practice for PLS-SEM~\citep{hair2019plssem}.
Table~\ref{tab:structural_results} presents the structural model results, including standardised path coefficients ($\beta$), standard errors, $t$-statistics, $p$-values, and 95\% bias-corrected confidence intervals.

\begin{table}[tb]
\small
\centering
\caption{Structural model results: path C
coefficients and hypothesis testing ($n = 154$)}
\label{tab:structural_results}
\begin{tabular}{llcccccl}
\hline
H & Path & $\beta$ & SE & $t$ & $p$ & 95\% CI & Result \\
\hline
H1 & PE $\rightarrow$ CUI & 0.476 & 0.055 & 8.677 & $<$.001 & [0.368, 0.582] & Supported \\
H2 & EE $\rightarrow$ CUI & 0.220 & 0.078 & 2.810 & .005 & [0.061, 0.372] & Supported \\
H3 & SI $\rightarrow$ CUI & \multicolumn{5}{c}{\textit{Excluded from model}} & Not tested \\
H4 & HM $\rightarrow$ CUI & 0.244 & 0.053 & 4.627 & $<$.001 & [0.141, 0.345] & Supported \\
H5 & FC $\rightarrow$ CUI & 0.042 & 0.083 & 0.505 & .614 & [$-$0.115, 0.208] & Not supp. \\
\hline
\multicolumn{8}{l}{\footnotesize \textit{Note:} Bootstrap: 5,000 subsamples.} \\
\end{tabular}
\end{table}

Three of four tested hypotheses were supported.
\PE emerged as the strongest predictor of \CUI ($\beta = 0.476$, $p < .001$), followed by \HM ($\beta = 0.244$, $p < .001$) and \EE ($\beta = 0.220$, $p = .005$).
\FC showed a positive but non-significant effect ($\beta = 0.042$, $p = .614$), with the confidence interval crossing zero.
H3 (\SI $\rightarrow$ \CUI) was not tested because the \SI construct was excluded from the model due to inadequate measurement properties, as detailed in Sect.~\ref{sec:phase2-measurement}.

\paragraph{Explanatory Power and Effect Sizes}

The structural model explained 64.7\% of variance in \CUI ($R^2 = 0.647$, adjusted $R^2 = 0.637$).
According to Cohen's guidelines as applied in PLS-SEM research~\citep{hair2019plssem}, this represents substantial explanatory power.
Table~\ref{tab:effect_sizes} presents the effect sizes ($f^2$) for each predictor, indicating their individual contribution to explaining variance in the endogenous construct.
\PE demonstrated a large effect ($f^2 = 0.368$), while \HM showed a small-to-medium effect ($f^2 = 0.116$) and \EE a small effect ($f^2 = 0.062$).
\FC showed a negligible effect ($f^2 = 0.002$).

\begin{table}[tb]
\small
\centering
\caption{Effect sizes ($f^2$) for predictors of Continued Use Intention}
\label{tab:effect_sizes}
\begin{tabular}{lcc}
\hline
Predictor & $f^2$ & Effect Size \\
\hline
Performance Expectancy & 0.368 & Large \\
Hedonic Motivation & 0.116 & Medium \\
Effort Expectancy & 0.062 & Small \\
Facilitating Conditions & 0.002 & Negligible \\
\hline
\multicolumn{3}{l}{\footnotesize \textit{Note:} Thresholds: 0.02 (small), 0.15 (medium), 0.35 (large)} \\
\vspace{-3mm}
\end{tabular}
\end{table}

\paragraph{Predictive Relevance}
We assessed the model's predictive relevance using PLSpredict with 10-fold cross-validation and 10 repetitions.
The model achieved $Q^2_{\text{predict}} = 0.625$, indicating substantial predictive relevance and strong out-of-sample predictive capability~\citep{shmueli2019predictive}.

\paragraph{Model Fit}
We computed the Standardised Root Mean Square Residual (SRMR) to quantify how well the estimated model reproduces the relationships observed in the data.
The SRMR for our model was 0.073, indicating good fit as values below 0.08 are considered acceptable~\citep{hair2019plssem}.
This result, combined with the substantial explanatory power ($R^{2} = 0.647$) and predictive relevance ($Q^{2}_{\text{predict}} = 0.625$), provides convergent evidence for our model's adequacy.

\paragraph{Model Justification}

Our decision to exclude \SI from the structural model was based on convergent evidence from multiple sources.
First, the EFA revealed that all three \SI items exhibited weak loadings (0.14--0.49) and high uniqueness values (0.73--0.90), indicating that \SI does not function as a coherent construct in this context.
Second, the Phase~1 pilot study at Apuliasoft showed similar measurement challenges with \SI.

In addition, FC4 was removed based on its outer loading (0.549), which fell below the recommended 0.708 threshold~\citep{hair2019plssem}.
After removing FC4, AVE increased from 0.667 to 0.807, and composite reliability improved from 0.885 to 0.926.

\paragraph{Structural Model Summary}

\begin{figure}[tb]
\small
\centering
\includegraphics[width=0.6\textwidth]{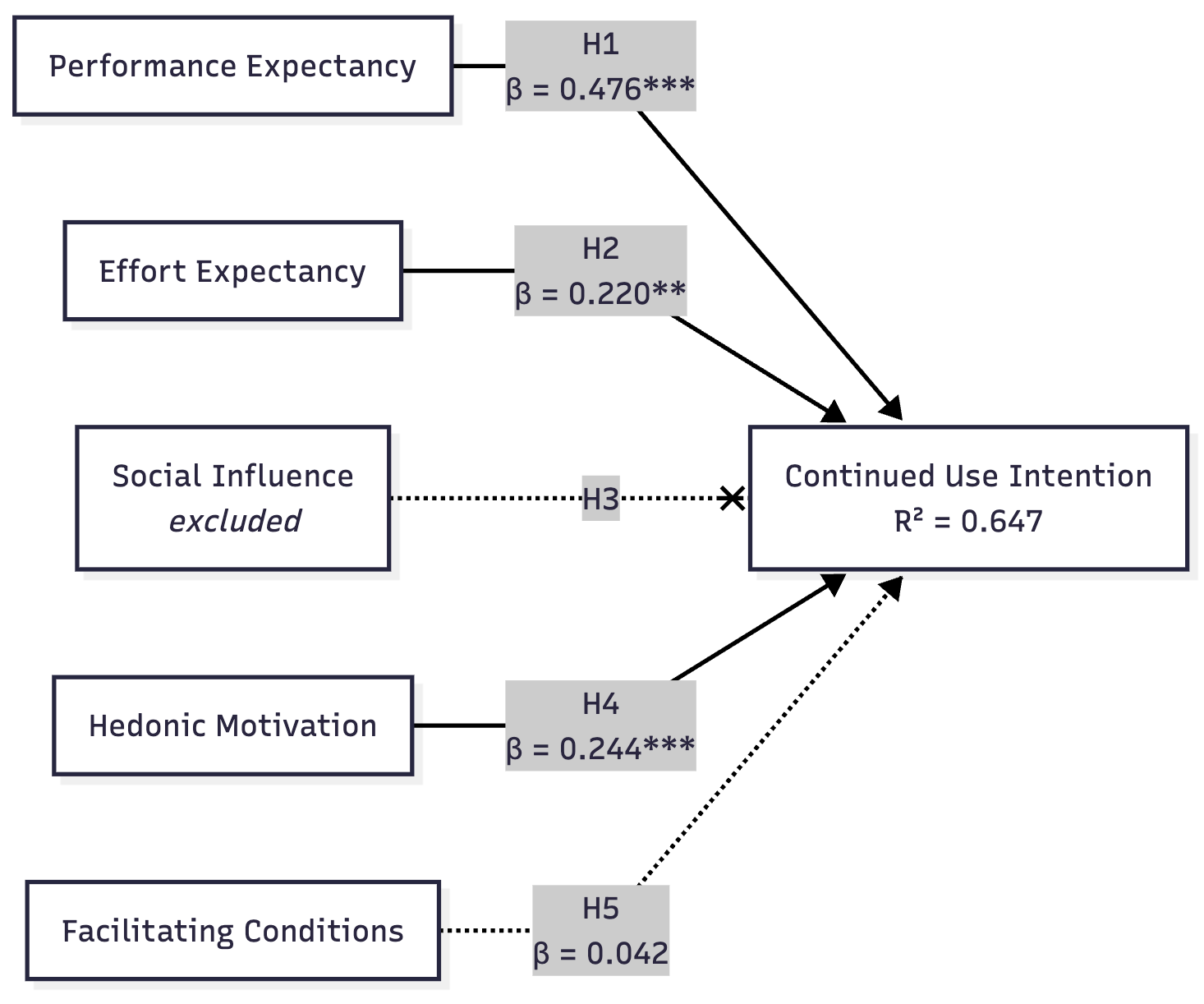}
\caption{Structural model results. Social Influence (H3) removed from the model. Path coefficients shown with significance levels: *** $p < .001$, ** $p < .01$. Dashed line indicates non-significant path.}
\vspace{-3mm}
\label{fig:structural_model}
\end{figure}

Figure~\ref{fig:structural_model} shows the structural model with path coefficients and significance levels.
In summary, the structural model demonstrates substantial explanatory power ($R^2 = 0.647$) and strong predictive relevance ($Q^2_{\text{predict}} = 0.625$).
Of the five hypotheses proposed, four were testable after excluding \SI due to measurement issues.
Three hypotheses were supported: H1 (\PE $\rightarrow$ \CUI), H2 (\EE $\rightarrow$ \CUI), and H4 (\HM $\rightarrow$ \CUI).
H5 (\FC $\rightarrow$ \CUI) was not supported.
\PE emerged as the dominant predictor of \CUI with a large effect size ($\beta = 0.476$, $f^2 = 0.368$), followed by \HM ($\beta = 0.244$, $f^2 = 0.116$) and \EE ($\beta = 0.220$, $f^2 = 0.062$).
\FC showed no significant effect ($\beta = 0.042$, $p = .614$).

\section{Discussion}\label{sec:discussion}

The results presented in the previous section reveal a consistent pattern across the two study phases. 
In the following, we discuss these findings and consider their implications.
Our key findings on the model factors and hypotheses noted in Sect.~\ref{sec:framework} are now highlighted in Sect.~\ref{sec:key-findings}, followed by implications for both organisations---especially SMEs---and future research (Sect.~\ref{sec:apuliasoft-mgmt-tranches} and \ref{sec:implications-research}, respectively).
In discussing the implications for organisations, we include the perspectives of Apuliasoft management, based on their experience with sustained GenAI use.

\subsection{Key Findings}\label{sec:key-findings}

Our study examined factors influencing software developers' intentions to continue using GenAI tools in Italian SMEs, applying an adapted UTAUT2 framework to the post-adoption context.
The PLS-SEM analysis revealed that three factors significantly predict continued use intention, with the model explaining 64.7\% of variance in \cui.
In contrast, \fc had no significant influence on \cui, and \si could not be tested due to persistent measurement failure across both study phases.
These findings suggest that in voluntary adoption contexts within SMEs, developers' continued use of GenAI tools depends primarily on perceived performance benefits and engagement, while organisational support plays a surprisingly limited role.

\paragraph{Performance as Dominant Predictor}

\pe emerged as the dominant predictor ($\beta = 0.476$, $f^2 = 0.368$), indicating that developers who perceive GenAI tools as enhancing their productivity and task accomplishment are substantially more likely to sustain their use.
This aligns with both meta-analytic evidence across technology adoption studies~\citep{kuttimani2021utaut2meta} and prior research in SE contexts~\citep{russo_navigating_2024}, which consistently identify \PE as the strongest factor predicting behavioural intention.
\citet{badghish2024sme} similarly found that perceived productivity benefits significantly predicted AI adoption among Saudi SMEs, consistent with the dominance of \PE in our model.
Our findings confirm that this pattern extends to the post-adoption phase.

The strength of this relationship is notable given our sample composition: 78.6\% of respondents had four or more years of professional experience and had already integrated GenAI into their workflows.
The dominance of \PE in a sample of experienced, regular GenAI users suggests that perceived performance benefits remain salient even after the initial adoption phase.

This finding is also consistent with Expectation-Confirmation Theory~\citep{bhattacherjee2001ect}, which posits that continued use is driven by users confirming pre-adoption expectations through actual experience.
The same finding is, instead, in contrast with~\citet{russo_navigating_2024}, who found that perceived usefulness did not directly predict initial GenAI adoption among software engineers, but rather influenced adoption indirectly through perceived workflow compatibility.
The divergence likely reflects the different adoption stages examined: at the initial adoption stage, developers may lack sufficient firsthand experience with a tool and thus assess its usefulness based on workflow compatibility; at sustained use, as noted in the Apuliasoft case study, developers have direct evidence of productivity gains and have learned to use GenAI effectively, thereby making performance perceptions more salient.
The Phase~1 longitudinal data illustrate this evolution directly: initial expectations about development speed were confirmed after six months of sustained use, while perceptions of software quality became more nuanced as developers encountered practical limitations such as hallucinations and context-inappropriate suggestions

\paragraph{The Role of Enjoyment}

\hm was the second-strongest predictor ($\beta = 0.244$, $f^2 = 0.116$), indicating that developers who find GenAI enjoyable are more likely to sustain their use.
This finding is noteworthy because professional software development is typically framed as utilitarian~\citep{beecham2008motivation,vanderheijden2004misq}, where intrinsic enjoyment might seem secondary to productivity.
Instead, we found that the hedonic dimension contributes meaningfully to continued engagement, beyond perceived performance benefits.
Consistently, a previous study on technology acceptance involving 268 design engineers found \HM to be the \emph{strongest} predictor of AI adoption intention~\citep{alharthi2025adoption}, providing further evidence that enjoyment matters even in technical professional contexts. In a software engineering context,~\citet{pereira_early_2024} found that a third of the developers studied reported increased enjoyment of programming when using GenAI. This perhaps indicates that the use of GenAI makes the work more enjoyable, rather than the tool itself.

The conversational, interactive nature of tools like ChatGPT may explain this effect.
Unlike traditional development tools that require learning specific syntax or navigating complex interfaces, GenAI enables creative experimentation through natural language dialogue. Developers can explore prompting strategies and engage in collaborative problem-solving with an AI assistant.
\citet{barke2023grounded} identified two distinct interaction modes in their grounded theory study of programmers using AI assistants: an \emph{acceleration mode} for familiar tasks, and an \emph{exploration mode} where developers deliberately prompt the AI to discover options, compare approaches, and engage in collaborative problem-solving.
This exploratory engagement, described by~\citet{ross2023programmer} as co-creative interaction, appears to generate satisfaction beyond mere task completion.
Research on conversational agents more broadly has found that users value entertainment and creative interaction alongside practical utility~\citep{skjuve2023user}, suggesting that conversational interfaces activate engagement mechanisms beyond those typically supported by traditional software.


\paragraph{Ease of Use Still Matters}
\ee also showed a significant positive effect ($\beta = 0.220$, $p = .005$), confirming that perceived ease of use influences sustained engagement even among experienced developers.
Although apparently counterintuitive given our population, GenAI tools present distinctive usability challenges that differ from traditional programming tools.
As found in the Apuliasoft case study, effective use requires developing prompt engineering skills, comprehending and critically evaluating AI-generated outputs---challenges that persist regardless of programming expertise.
\citet{vaithilingam2022expectation} found that developers faced significant difficulties understanding, editing, and debugging Copilot-generated code, which hindered task-solving effectiveness despite overall positive perceptions of the tool.
Similarly,~\citet{he2025empirical} identified prompt design as a key challenge category in their analysis of LLM developer questions, noting that developers must iteratively craft and refine prompts to achieve satisfactory outputs---a skill that does not transfer from conventional programming experience.
Our finding suggests that when developers perceive these novel interaction patterns as manageable, they are more inclined toward continued use.

\paragraph{Limited Role of Organisational Support}
Contrary to our hypothesis, \fc showed no significant effect ($\beta = 0.042$, $p = .614$).
This null finding aligns with previous GenAI adoption studies in professional settings. 
\citet{kim2024determinants} found \FC non-significant among employees in Korean companies, 
and \citet{lambiase_investigating_2025} report similar results for software engineers adopting LLMs.
Two contextual factors may further explain this pattern.
First, GenAI tools are predominantly self-service technologies.
Unlike enterprise systems dependent on IT support and training programmes, tools like ChatGPT can be accessed directly with minimal organisational facilitation.
The high mean score on \FC items ($M = 4.02$) suggests most developers already perceive adequate support.
Second, voluntary adoption may create a selection effect: developers who lack necessary resources or encounter compatibility issues may simply not adopt, while those who do adopt have already overcome potential barriers.
\FC may thus function as a threshold condition for initial adoption rather than a driver of sustained use.

\paragraph{Measurement Challenges with Social Influence}
\si was excluded from our model due to persistent measurement problems: all three items exhibited weak loadings (0.14--0.49) and high uniqueness values (0.73--0.90) in the Phase~2 exploratory factor analysis.

This pattern likely reflects the voluntary adoption context rather than a methodological limitation.
The standard \SI items from UTAUT2 ask whether colleagues, managers, or other important people think one should use a technology, but such social expectations may carry little weight when developers independently choose their tools.
In mandatory contexts, management directives and colleague compliance create pressure to adopt. In voluntary settings like ours---as confirmed by the sample's high voluntariness scores 
(see Sect.~\ref{sec:sample-characteristics})---adoption decisions rest primarily on individual assessment.
This interpretation aligns with \citet{venkatesh2003utaut}, who found that \SI is more influential in mandatory contexts and its effect diminishes with experience.
\citet{russo_navigating_2024} similarly found that social factors did not significantly predict intention to use LLMs among software engineers.
Consistently, the Phase~1 case study found that GenAI was used almost exclusively for individual tasks, with minimal adoption in collaborative settings such as team meetings, further suggesting that social dynamics play a limited role in shaping continued use decisions.
Overall, these findings suggest that \SI matters more for mandatory technology rollouts than when (experienced) professionals choose to adopt tools voluntarily.




\subsection{Implications for organisations}\label{sec:apuliasoft-mgmt-tranches}

Based on the findings from the two phases and the resulting discussion of these with Apuliasoft management, we provide several implications for SMEs to support continued use of GenAI tools beyond initial adoption. These recommendations acknowledge the constraints (e.g., financial, human capital) that SMEs operate under, as well as the greater freedom they have compared to larger firms, especially in terms of experimentation before early standardization of tooling.

\paragraph{Frame GenAI as one tool among many}
The most emphatic message from Apuliasoft management was straightforward: treat GenAI as another instrument in the development toolchain.
This framing carries important implications.
It normalises adoption by removing the sense that GenAI use is exceptional or threatening, and it establishes clear expectations.
Just as developers are responsible for the code they write with any tool, they remain accountable for GenAI-assisted output.
Management reported encountering cases in which developers attributed documentation errors to the AI, thereby disclaiming responsibility.
The CEO's response was direct: ``\textit{regardless of the tool used, you own your work.}''
SMEs adopting GenAI should establish this principle early, treating AI-generated output as raw material requiring the same quality standards and review processes as any other development artefact.

\paragraph{Allow experimentation, then standardise}
Our finding that \hm significantly predicts continued use surprised Apuliasoft management at first, as they had not considered the enjoyment dimension before.
Yet upon reflection, management recognised that developers who found GenAI engaging were indeed its most effective users, including one architect who pays for Cursor IDE from his own pocket and has become an internal evangelist after he completed a major architectural redesign in two days, estimating that without AI assistance, the work would have required three weeks. Similarly, a junior developer assigned to prototype an application finished in a few days, rather than the month typically expected for new hires at a similar career stage. 
This evidence, while anecdotal, illustrates the productivity perceptions that our quantitative findings identify as key to sustained use.
SMEs should consider a staged approach to tool selection: initially, allow developers to experiment freely with various tools to discover which work best for their specific needs and working styles.
Once patterns emerge, organisations can standardise on specific tools and practices.
Apuliasoft has since adopted Gemini Pro company-wide---chosen for integration with their existing Google ecosystem---while maintaining voluntary usage and with higher usage tiers based on seniority, to avoid over-reliance in younger employees.
The key insight is that imposed standardisation without prior experimentation would suppress precisely the intrinsic engagement that anticipates sustained use.

\paragraph{Invest in methodology, not tool training}
Our quantitative finding that \fc had no significant effect on continued use aligns with management's experience. 
They confirmed that developers required no formal training to use GenAI tools effectively, since the natural language interface eliminates traditional learning curves. 
This view contrasts with prior GenAI studies that recommend training (e.g.,~\cite{kemell_still_2025, lambiase_investigating_2025}).
However, they emphasised that \textit{methodological training} with GenAI tools is instead essential. 
Developers need guidance not on how to prompt an AI, but on how AI-assisted development integrates with existing quality standards, review processes, and development methodology. 
Apuliasoft is developing an internal AI policy that addresses questions such as when `vibe coding' is permitted (only for experimentation and rapid prototyping), what review processes apply to AI-generated code, and how to maintain documentation standards when AI assists with technical writing.
This methodological focus becomes more pressing as GenAI capabilities expand. 
The COO expressed concern about the increasing proliferation of new tools, such as agentic AI systems that can execute multi-step workflows autonomously and produce cascading effects across codebases, making process integration and quality controls substantially harder. 
SMEs should therefore redirect training investments from prompt engineering---which management believes is diminishing in importance---toward integrating GenAI into existing quality assurance frameworks, while also establishing mechanisms for ongoing reassessment as tool capabilities evolve and new tools emerge.

\paragraph{Abandon self-managed GenAI infrastructure aspirations}
Finally, one striking lesson from Apuliasoft's experience addresses a concern many companies (including SMEs) share (e.g., ~\cite{kemell_still_2025}, \cite{davila_industry_2024}): anxiety about sending code to third-party services.
Motivated by confidentiality concerns when working with client intellectual property, management explored alternatives to commercial GenAI APIs throughout the study period.
On-premise deployment was ruled out immediately as economically unviable.
To test self-hosted cloud as a middle ground, the team ran a week-long trial of a heavily quantised and distilled open-weight model on rented GPU instances.
Monthly costs soon rose into the thousands of euros, while response quality and latency were poor enough that developers abandoned the experiment within days, reverting to commercial services.
For a full-size, minimally compressed model, they estimated costs in the tens to hundreds of thousands of euros per month, far exceeding what an SME could justify. The implication for SMEs is that self-managed infrastructure is not a viable path, and the real challenge lies in navigating the legal and contractual dimensions of cloud-based AI services.
Apuliasoft continues to navigate this still uncharted territory, working with legal counsel to determine when GenAI use is permissible for consulting, though the CEO noted with frustration that the same code and documents routinely pass through cloud-hosted repositories and file-sharing services without comparable scrutiny.

\subsection{Implications for Research}\label{sec:implications-research}

\paragraph{Natural language interaction as an explanatory lens}

The pattern of effects we observed may be partially explained by a fundamental shift in human-computer interaction: GenAI tools are the first widely adopted professional productivity tools accessed primarily through natural language.
Unlike traditional development environments that require mastering IDE conventions, command syntax, or API structures, conversational AI allows developers to express intent directly.
This interaction modality appears to lower traditional usability barriers; \citet{liang_large-scale_2024} found that developers value AI programming assistants precisely because they reduce keystrokes and help recall syntax, suggesting that ease-of-use concerns become less differentiating once users can communicate in natural language.
At the same time, conversational interaction may amplify hedonic engagement by inviting exploration and creative problem-solving in ways that transactional interfaces do not \citep{Sun_GenAI_Wild_24}.
Our findings suggest that current acceptance models may need updating for conversational AI tools: future research should investigate whether natural language interfaces systematically influence technology acceptance dynamics in professional contexts.

\vspace{-2mm}
\paragraph{Hedonic motivation and utilitarian assumptions}
The significance of \hm as the second-strongest predictor challenges utilitarian assumptions that have shaped technology acceptance research in software engineering contexts.
Studies of professional technology acceptance have typically emphasised performance and effort constructs~\citep{davis1989tam,venkatesh2003utaut}, a pattern confirmed in software developer contexts~\citep{riemenschneider2002explaining}, with meta-analytic evidence suggesting that hedonic effects are weaker in utilitarian contexts~\citep{tamilmani2019ijinfoman-battle}.
Our findings suggest this assumption may not hold for GenAI tools.
Whether \HM's importance reflects the novelty of these tools, their distinctive conversational modality, or broader characteristics of creative professional work remains an open question.
Future research should track whether hedonic factors retain predictive power as GenAI tools mature and become a commodity, which would help distinguish novelty effects from more durable features of natural language interfaces, and examine contexts such as OSS development, where intrinsic motivation is recognised as a primary driver of contribution~\citep{gerosa2021icse-shifting}.

\vspace{-2mm}
\paragraph{Temporal dynamics in adoption predictors}
Our findings diverge from~\citet{russo_navigating_2024}, who found that workflow compatibility drove initial adoption intention among software engineers.
In our continued-use context, by contrast, \pe emerged as the dominant predictor.
This pattern suggests that the factors shaping adoption decisions shift across the technology lifecycle.
At initial adoption, when users lack firsthand evidence of a tool's capabilities, contextual factors such as workflow fit and organisational support may matter more.
At sustained use, accumulated productivity evidence becomes the primary consideration.
This pattern is consistent with Expectation-Confirmation Theory~\citep{bhattacherjee2001ect}, which posits that post-adoption behaviour is shaped by confirmed experience rather than anticipated benefits.
Adoption models should therefore specify which lifecycle stage they address, rather than assuming the same predictors matter equally from initial trial through established use.
Longitudinal designs tracking users from first exposure onward would help clarify how these relationships change over time.

\vspace{-2mm}
\paragraph{Extension of UTAUT2 to post-adoption contexts}
At the model level, the substantial explanatory power ($R^2 = 0.647$) confirms that UTAUT2 can be meaningfully adapted for continued use intentions, not just initial adoption, providing a foundation for future post-adoption studies.
However, not all UTAUT2 constructs transfer equally well to voluntary professional contexts.
As noted above, \si exhibited persistent measurement failure, suggesting that traditional social pressure items may require either adaptation or explicit acknowledgement of limited applicability when studying professional tool adoption in non-mandatory settings.
At the same time, the remaining unexplained variance suggests that other constructs play a role in sustained use decisions.
Trust in the reliability and correctness of AI-generated output is one such factor.
While \pe captures perceived productivity rather than confidence in output quality, trust addresses the persistent concerns about inconsistent suggestions~\citep{mastropaolo2023robustness} and the cognitive burden of evaluating them~\citep{liang_large-scale_2024}.
Similarly, as GenAI use matures beyond the early post-adoption phase studied here, \hab may gain explanatory relevance, consistent with evidence that continued use becomes increasingly automatic over time~\citep{limayem2007habit}.
Future research should consider extending our model with these and related constructs, such as task-specific fit across different GenAI use cases, to determine whether they account for additional variance in \cui.

\section{Threats to Validity}\label{sec:threats}

\subsection{Internal Validity}
Phase~1 faced operational challenges from prevalent remote work at Apuliasoft, limiting opportunities for naturalistic observation. 
We addressed this by scheduling dedicated interview sessions.
 Additionally, one researcher had prior familiarity with Apuliasoft through an earlier internship, which may have influenced participant responses. We mitigate the risk of social desirability bias by ensuring survey anonymity and triangulating interview data with survey responses and ethnographic observations.

Phase~2 relies on self-reported survey data collected at a single time point. 
Because all constructs were measured using the same instrument administered simultaneously, common method bias represents a potential threat. 
We implemented procedural safeguards (respondent anonymity) and conducted statistical tests: Harman's single-factor test yielded 40.23\% variance explained by the first factor, below the 50\% threshold.
While this test has known limitations as a definitive CMB assessment~\citep{podsakoff2003cmb}, the result, combined with the absence of inter-construct correlations exceeding 0.90, provides evidence that CMB does not pose a serious threat, though single-source bias cannot be ruled out entirely.

The Phase~2 cross-sectional design limits causal inference between perceptions and intentions. 
Although Phase~1 provides supporting temporal evidence through its six-month longitudinal component, we cannot isolate effects associated with the initial adoption decision since all Phase~1 participants were already GenAI users at baseline. 
In addition, Phase~1 sample attrition (37\%) may have introduced survival bias if developers who discontinued GenAI use were less likely to complete the follow-up survey.

\subsection{Construct Validity}
Our study measures continued use intention rather than actual usage behaviour. 
While behavioural intention predicts subsequent behaviour with moderate strength~\citep{sheeran2002intention,venkatesh2003utaut}, a gap between intention and action may persist. 
This concern is partly mitigated by our research context: all respondents were already active GenAI users and, as such, continued use intention reflects the decision to maintain an established behaviour rather than initiate a new one. 
Our Phase~1 longitudinal data, where we triangulated stated intentions with self-reported and observed usage patterns, found reasonable alignment within the SME setting. Nonetheless, objective metrics such as IDE telemetry would strengthen conclusions about actual usage.

We adapted the UTAUT2 scales to the GenAI context, drawing on recent work in software development~\citep{vaz_pereira_exploring_2025}. 
The Phase~1 pilot provided an opportunity to test item comprehension and refine Italian translations before confirmatory deployment in Phase~2.
Within the PLS-SEM framework, all retained constructs met thresholds for average variance extracted (AVE $\geq 0.50$) and the heterotrait-monotrait ratio (HTMT $< 0.85$), supporting measurement model validity~\citep{hair2019plssem}.

The \SI construct was excluded from the final model due to poor measurement properties. 
Our exploratory factor analysis revealed a two-factor structure rather than the intended six factors: Factor~1 captured \PE, \HM, and \CUI items (outcome-oriented aspects), whereas Factor~2 captured \EE and \FC items (ease and support dimensions). 
All three \SI items exhibited weak loadings and high uniqueness values (0.73--0.90), suggesting that \SI may not function as a coherent construct in voluntary GenAI adoption among experienced developers. 
This limits our ability to draw conclusions about \SI effects.

Finally, our model focuses on five UTAUT2 constructs, omitting factors such as trust in AI-generated output and perceived risks~\citep{russo_navigating_2024}. 
Our Phase~1 qualitative data corroborated that such concerns exist among developers. 
Future research should integrate these constructs for more comprehensive models of GenAI adoption.

\subsection{External Validity}
Our Phase~2 sampling strategy relied on convenience sampling through professional networks and social media~\citep{baltes2022sampling}. 
Developers who voluntarily participate in GenAI research may be more enthusiastic adopters than the broader SME population, potentially inflating positive perceptions. 
Because the survey was distributed through channels with unknown reach, we cannot calculate a meaningful response rate, which limits our ability to assess non-response bias.
Similarly, we did not collect data on industry sector, so we cannot determine whether certain domains are over- or under-represented.

The alignment between Phase~1 findings (situated at a single Italian SME) and Phase~2 findings (across multiple Italian SMEs) suggests our results generalise to the broader population of Italian SME developers, and possibly to SMEs in other countries given similar resource constraints. 
Our findings also broadly align with GenAI adoption studies in other contexts, including \citet{vaz_pereira_exploring_2025} at a large Brazilian organization, \citet{kim2024determinants} among employees in Korean companies, and \citet{badghish2024sme} in Saudi SMEs, suggesting the patterns we observed are not unique to the Italian context.
However, generalisability to larger enterprises remains uncertain: large organisations have greater capacity to invest in training and support structures, which could alter the influence of contextual factors on adoption outcomes.

Both phases examined voluntary GenAI adoption, so findings may not generalise to mandatory implementation settings. Additionally, our sample consists exclusively of employees who perform development-related activities (e.g., designing, coding, code reviewing, testing); therefore, findings may not extend to other roles involved in software development, such as product owners and project managers. 
Finally, GenAI represents a rapidly evolving domain. We collected data between Aug. 2024 and Sept. 2025; subsequent developments may alter the factors driving continued use.

\section{Conclusion}\label{sec:conclusion}

This study investigated factors sustaining software developers' continued use of GenAI, addressing a gap in research that has predominantly examined initial adoption.
Through a six-month longitudinal pilot and a cross-sectional survey of 154 developers across Italian SMEs, we developed and validated a UTAUT2-based model explaining 64.7\% of variance in continued use intention.

\pe emerged as the dominant predictor, confirming that tangible productivity gains drive sustained engagement.
The significant contribution of \hm was less expected, suggesting that GenAI's conversational interface fosters intrinsic enjoyment even in professional contexts.
Notably, organisational support showed no significant effect and \si exhibited persistent measurement failure, raising an open question for future research: whether continued use decisions in voluntary professional contexts are inherently individual and instrumental, or whether the standard UTAUT2 items require adaptation to capture how social factors operate in post-adoption settings.

These findings indicate that post-adoption dynamics differ substantively from initial adoption.
For SMEs seeking to sustain GenAI use, evident performance benefits matter most, but the engagement these tools can foster should not be overlooked.

\begin{acks}
We thank Apuliasoft for their participation in this study. 
We are also grateful to the anonymous survey participants.
Alexandra Pajonk was a visiting student at the University of Bari during the study period.
This paper was supported by the Ministry of Science, Technology, and Innovation of Brazil (Law 8.248 from Oct 23, 1991), within the scope of PPI-SOFTEX, coordinated by Softex, and published in the Residência em TIC 02 - Aditivo, Official Gazette 01245.012095/2020-56. Guilherme Pereira is supported by the Federal Institute of Education, Science and Technology of Rio Grande do Sul (IFRS). This study was financed in part by the Coordenação de Aperfeiçoamento de Pessoal de Nível Superior - Brasil (CAPES) - Finance Code 001.
\end{acks}

\bibliographystyle{elsarticle-harv} 
\bibliography{refs}

\newpage
\appendix
\section{Ethics}\label{app:ethics}
The study received approval from the University of Bari's Research Ethics Committee (no. CER\_19A77AFDB25). This approval covered both the Phase 1 longitudinal case study at Apuliasoft and the Phase 2 cross-sectional survey across multiple Italian SMEs.
For Phase 1, the CEO and COO of Apuliasoft were briefed on the study protocol and provided organizational consent for researcher access to employees. Participation remained entirely voluntary for all Apuliasoft developers, with no pressure from management to participate. All surveys were completed anonymously, and data from interviews and observations were anonymized. Results shared with company management were presented in aggregate form only, ensuring individual employee responses could not be identified.
For Phase 2, all recruitment communications emphasized the study's academic nature, voluntary participation, complete anonymity of responses, and approximate completion time ($\sim$10 minutes). Participation was entirely voluntary with no monetary compensation or incentives offered.
Across both phases, all participants provided informed consent before participating, with clear information about the study purposes, data handling procedures, guarantees of anonymity, and their right to withdraw at any point. No personally identifiable information was collected, and all responses remained accessible only to the research team.

\newpage
\section{S2 Survey Items and Correlations}
\label{app:phase1}

\begin{table}[!htbp]
\small
\centering
\caption{Measurement items by construct, adapted from the original UTAUT2 scale~\citep{venkatesh2012utaut2}}
\label{tab:survey_items}
\footnotesize
\begin{tabular}{lp{11cm}}
\hline
Item & Statement \\
\hline
\multicolumn{2}{l}{\textit{Performance Expectancy (PE)}} \\
PE1 & I find GenAI tools useful in my daily work \\
PE2 & Using GenAI tools increases my chances of achieving things that are important to me \\
PE3 & Using GenAI tools helps me accomplish things more quickly \\
PE4 & Using GenAI tools increases my productivity \\
\hline
\multicolumn{2}{l}{\textit{Effort Expectancy (EE)}} \\
EE1 & Learning how to use GenAI tools is easy for me \\
EE2 & My interaction with GenAI tools is clear and understandable \\
EE3 & I find GenAI tools easy to use \\
EE4 & It is easy for me to become skilful at using GenAI tools \\
\hline
\multicolumn{2}{l}{\textit{Social Influence (SI)}$^\dagger$} \\
SI1 & People who are important to me think that I should use GenAI tools \\
SI2 & People who influence my behaviour (e.g., team leads, managers, mentors) think that I should use GenAI tools \\
SI3 & People whose opinions I value prefer that I use GenAI tools \\
\hline
\multicolumn{2}{l}{\textit{Hedonic Motivation (HM)}} \\
HM1 & Using GenAI tools is fun \\
HM2 & I enjoy using GenAI tools \\
HM3 & I find using GenAI tools satisfying \\
\hline
\multicolumn{2}{l}{\textit{Facilitating Conditions (FC)}$^\ddagger$} \\
FC1 & I have the resources necessary to use GenAI tools \\
FC2 & I have the knowledge necessary to use GenAI tools \\
FC3 & GenAI tools are compatible with other technologies I use \\
FC4 & I can get help from others when I have difficulties using GenAI tools \\
\hline
\multicolumn{2}{l}{\textit{Continued Use Intention (CUI)}} \\
CUI1 & I intend to continue using GenAI tools in the future \\
CUI2 & I will always try to use GenAI tools in my daily work \\
CUI3 & I plan to use GenAI tools regularly \\
\hline
\multicolumn{2}{p{11cm}}{\footnotesize $^\dagger$SI construct excluded from final model.} \\
\multicolumn{2}{p{11cm}}{\footnotesize $^\ddagger$FC4 item excluded from final model.} \\
\multicolumn{2}{p{11cm}}{\footnotesize \textit{Note.} Items were administered in Italian (English translations are provided for reference) and measured on 5-point Likert scales (1 = Strongly Disagree to 5 = Strongly Agree). The complete survey instrument is available in the supplementary material.} \\
\end{tabular}
\end{table}

\begin{table}[!h]
\small
\centering
\caption{Correlation matrix of UTAUT2 constructs from survey S2 (n=17).}
\label{tab:correlation_matrix}
\begin{tabular}{lcccccccc}
\hline
 & PE & EE & SI & FC & HM & PV & H & CUI \\
\hline
Performance Expectancy & 1.00 & & & & & & & \\
Effort Expectancy & 0.13 & 1.00 & & & & & & \\
Social Influence & $-$0.11 & 0.50* & 1.00 & & & & & \\
Facilitating Conditions & 0.24 & 0.08 & $-$0.19 & 1.00 & & & & \\
Hedonic Motivation & 0.28 & $-$0.06 & 0.26 & 0.12 & 1.00 & & & \\
Price Value & 0.18 & 0.21 & 0.15 & 0.09 & 0.31 & 1.00 & & \\
Habit & 0.22 & 0.12 & 0.19 & 0.14 & 0.25 & 0.28 & 1.00 & \\
Continued Use Intention & 0.30 & $-$0.03 & 0.13 & 0.15 & $-$0.08 & 0.05 & 0.17 & 1.00 \\
\hline
\multicolumn{7}{l}{\footnotesize *$p < .05$}
\end{tabular}
\end{table}

\newpage
\section{S3 Survey Exploratory Factor Analysis}\label{app:phase2}

\begin{table}[!h]
\small
\centering
\caption{Exploratory factor analysis with oblimin rotation ($n$ = 149)}
\label{tab:efa_pattern}
\footnotesize
\begin{tabular}{llrrrc}
\hline
Item & Construct & Factor 1 & Factor 2 & Unique. & Status \\
\hline
PE1  & PE  & \textbf{0.67} &  0.19 & 0.52 & OK \\
PE2  & PE  & \textbf{0.83} & $-$0.03 & 0.31 & OK \\
PE3  & PE  & \textbf{0.83} & $-$0.02 & 0.31 & OK \\
PE4  & PE  & \textbf{0.85} & $-$0.09 & 0.28 & OK \\
\hline
EE1  & EE  & $-$0.04 & \textbf{0.79} & 0.37 & OK \\
EE2  & EE  &  0.02 & \textbf{0.86} & 0.26 & OK \\
EE3  & EE  & $-$0.15 & \textbf{0.85} & 0.26 & OK \\
EE4  & EE  &  0.10 & \textbf{0.76} & 0.42 & OK \\
\hline
SI1  & SI  &  0.14 &  0.29 & 0.90 & Weak \\
SI2  & SI  & \textit{0.37} &  0.12 & 0.85 & Weak \\
SI3  & SI  & \textit{0.49} & $-$0.17 & 0.73 & Marginal \\
\hline
FC1  & FC  &  0.01 & \textbf{0.77} & 0.40 & OK \\
FC2  & FC  &  0.12 & \textbf{0.79} & 0.37 & OK \\
FC3  & FC  &  0.27 & \textbf{0.65} & 0.51 & OK \\
FC4  & FC  & $-$0.10 & \textbf{0.57} & 0.67 & High uniq. \\
\hline
HM1  & HM  & \textbf{0.66} &  0.03 & 0.56 & OK \\
HM2  & HM  & \textbf{0.71} & $-$0.07 & 0.49 & OK \\
HM3  & HM  & \textbf{0.58} & $-$0.12 & 0.64 & High uniq. \\
\hline
CUI1 & CUI & \textbf{0.66} &  0.27 & 0.49 & OK \\
CUI2 & CUI & \textbf{0.76} &  0.06 & 0.41 & OK \\
CUI3 & CUI & \textbf{0.75} &  0.12 & 0.43 & OK \\
\hline
\multicolumn{2}{l}{SS Loadings} & 5.92 & 4.91 & & \\
\multicolumn{2}{l}{\% Variance} & 28.2 & 23.4 & & \\
\multicolumn{2}{l}{Cumulative \%} & 28.2 & 51.6 & & \\
\hline
\end{tabular}
\vspace{2mm}
\parbox{0.95\textwidth}{\footnotesize
\textit{Note.} Extraction method: Principal axis factoring. Rotation: Oblimin with Kaiser normalisation. 
Factor loadings $\geq 0.50$ in \textbf{bold}; loadings 0.30--0.49 in \textit{italics}.
Factor 1 primarily captures PE, HM, and CUI items; Factor 2 captures EE and FC items.
SI items show weak loadings and high uniqueness across both factors.
Factor correlation: $r$ = 0.57.
KMO = 0.875; Bartlett's test: $\chi^2$(210) = 2176.42, $p < .001$.
}
\end{table}

\section{Measurement Model}\label{app:measurement-model}

\begin{table}[!h]
\small
\centering
\caption{Measurement model: reliability and convergent validity ($n = 154$)}
\label{tab:measurement_model}
\begin{tabular}{lccccp{4cm}}
\hline
Construct & $\alpha$ & CR & AVE & Items & Loadings \\
\hline
Perform. Expect. & 0.904 & 0.933 & 0.778 & 4 & 0.81, 0.92, 0.91, 0.88 \\
Effort Expect. & 0.879 & 0.917 & 0.735 & 4 & 0.87, 0.91, 0.77, 0.87 \\
Hedonic Motiv. & 0.780 & 0.869 & 0.690 & 3 & 0.90, 0.85, 0.73 \\
Facilit. Condit. & 0.882 & 0.926 & 0.807 & 3 & 0.85, 0.93, 0.91 \\
Conti. Use Inten. & 0.899 & 0.937 & 0.832 & 3 & 0.88, 0.93, 0.93 \\
\hline
\multicolumn{6}{p{0.95\textwidth}}{\footnotesize \textit{Note.} $\alpha$ = Cronbach's alpha; CR = Composite Reliability; AVE = Average Variance Extracted. Internal consistency requires $\alpha \geq 0.70$ and CR $\geq 0.70$. Convergent validity requires AVE $\geq 0.50$ and loadings $\geq 0.708$; items with loadings $\geq 0.40$ may be retained if theoretically justified~\citep{russo_pls-sem_2021}.} \\
\end{tabular}
\end{table}

\begin{table}[!h]
\small
\centering
\caption{Discriminant validity: Heterotrait-Monotrait Ratio (HTMT)}
\label{tab:htmt}
\begin{tabular}{lccccc}
\hline
 & PE & EE & HM & FC & CUI \\
\hline
Performance Expectancy & -- &  &  &  &  \\
Effort Expectancy & 0.551 & -- &  &  &  \\
Hedonic Motivation & 0.617 & 0.400 & -- &  &  \\
Facilitating Conditions & 0.625 & 0.827 & 0.530 & -- &  \\
Continued Use Intention & 0.814 & 0.636 & 0.679 & 0.637 & -- \\
\hline
\multicolumn{6}{p{0.95\textwidth}}{\footnotesize \textit{Note.} HTMT compares correlations between indicators of different constructs to correlations within constructs. Values below 0.85 establish discriminant validity~\citep{henseler2015htmt}. The highest observed value (0.827, between EE and FC) remains below this threshold.} \\
\end{tabular}
\end{table}

\begin{table}[!h]
\small
\centering
\caption{Discriminant validity: Fornell-Larcker criterion}
\label{tab:fornell_larcker}
\begin{tabular}{lccccc}
\hline
 & PE & EE & HM & FC & CUI \\
\hline
Performance Expectancy & \textbf{0.882} &  &  &  &  \\
Effort Expectancy & 0.500 & \textbf{0.857} &  &  &  \\
Hedonic Motivation & 0.525 & 0.345 & \textbf{0.831} &  &  \\
Facilitating Conditions & 0.569 & 0.732 & 0.455 & \textbf{0.898} &  \\
Continued Use Intention & 0.739 & 0.573 & 0.590 & 0.585 & \textbf{0.912} \\
\hline
\multicolumn{6}{p{0.95\textwidth}}{\footnotesize \textit{Note.} Diagonal values (bold) = $\sqrt{\text{AVE}}$; off-diagonal = inter-construct correlations. Discriminant validity is established when $\sqrt{\text{AVE}}$ exceeds all correlations in its row and column; all constructs satisfy this criterion.} \\
\end{tabular}
\end{table}

\end{document}